\documentclass[aps,amsmath,twocolumn,amssymb,floatfixng,showpacs,
superscriptaddress,footinbib, longbibliography]{revtex4-1}

\usepackage{graphicx}
\usepackage{dcolumn}
\usepackage{multirow}
\usepackage{booktabs}
\usepackage{bm,color}
\usepackage{braket}
\usepackage{amsmath,amssymb}
\usepackage[colorlinks,linkcolor=blue,hyperindex,CJKbookmarks]{hyperref}
\usepackage{epstopdf}

\definecolor{red1}{rgb}{0.6,0,0}

\newcommand{\U}{\mathop{\rm {}U}}

\def\rb{\textcolor{blue}}
\newcommand{\bgea}{\begin{equation}}
\newcommand{\enea}{\end{equation}}
\newcommand{\bea}{\begin{eqnarray}}
\newcommand{\eea}{\end{eqnarray}}
\newcommand{\braOket}[3]{\langle n(#1)|#2|n(#3)\rangle}
\newcommand{\hatmu}{\hat\mu}

\begin{document}
\title{Topological magnon in exchange frustration driven incommensurate spin spiral of a kagome lattice YMn$_6$Sn$_6$}
\author{Banasree Sadhukhan}
\affiliation{Department of Physics and Nanotechnology, SRM Institute of Science and Technology, Kattankulathur, 603203, Chennai, Tamil Nadu, India}
\email{banasres@srmist.edu.in}

\affiliation{Department of Applied Physics, School of Engineering Sciences, KTH Royal Institute of Technology, AlbaNova University Center, SE-10691 Stockholm, Sweden}
\email{banasree@kth.se}
\author{Anders Bergman}
\affiliation{Department of Physics and Astronomy, Uppsala University, Box 516, SE-75120 Uppsala, Sweden}
\author{Patrik Thunstr{\"o}m}
\affiliation{Department of Physics and Astronomy, Uppsala University, Box 516, SE-75120 Uppsala, Sweden}
\author{Manuel Pereiro Lopez}
\affiliation{Department of Physics and Astronomy, Uppsala University, Box 516, SE-75120 Uppsala, Sweden}
\author{Olle Eriksson}
\affiliation{Department of Physics and Astronomy, Uppsala University, Box 516, SE-75120 Uppsala, Sweden}
\author{Anna Delin}
\affiliation{Department of Applied Physics, School of Engineering Sciences, KTH Royal Institute of Technology,  AlbaNova University Center, SE-10691 Stockholm, Sweden}
\affiliation{Swedish e-Science Research Center (SeRC), KTH Royal Institute of Technology, SE-10044 Stockholm, Sweden}
\affiliation{Wallenberg Initiative Materials Science for Sustainability (WISE), KTH Royal Institute of Technology, SE-10044 Stockholm, Sweden}
%


\begin{abstract}

YMn$_6$Sn$_6$ consists of two types of Mn-based kagome planes stacked along $c$-axis having a complex magnetic interactions.  We report a spin reconstruction in YMn$_6$Sn$_6$ from ferromagnet (FM) into a combination of two incommensurate spin spirals (SSs) originating from two different type of Mn kagome planes driven by frustrated magnetic exchanges along the $c$-axis with inclusion of Hubbard U.  The pitch angle and wave vector of the incommensurate SSs are $\sim$89.3$^{\circ}$ and  $\sim$ (0 0 0.248) respectively which are in excellent agreement with experiment.  We employed an effective model Hamiltonian constructed out of exchange interactions to capture experimentally observed non-equivalent nature of the two incommensurate SSs which also explain FM-SS crossover due to antiferromagntic spin exchange with correlation.  We further report the existence of topological magnon with spin-orbit coupling in incommensurate SS phase of YMn$_6$Sn$_6$ by calculating the topological invariants and Berry curvature profile.  The location of Dirac magnon in energy landscape at 73 meV matches with another experimental report.  We demonstrate the accuracy of our results by highlighting experimental features in YMn$_6$Sn$_6$.

\end{abstract}

\maketitle

{\it{\rb {Introduction :}}} Magnetic kagome crystal, consisting of corner-sharing triangles, is an intriguing class of materials in condensed-matter physics because of having Chern gapped topological fermions,  electronic flat bands,  topological magnon, non-trivial Hall effect and non-trivial topology \cite{Han2012,  PhysRevB.62.R6065,  Ghimire2020-qs,  RevModPhys.89.025003,  Han2012-ry,  Yin2019,  Liu2018,  Kang2020}.  The co-existence of geometric frustration and strong electron-electron correlations makes kagome-lattice based systems an ideal system to study the interplay between magnetism and topology for next-generation spintronics \cite{Ghimire2020-qs}.  Flat bands, with the quenching kinetic energy,  appear in kagome magnets due to lattice  geometry and hosts superconductivity \cite{Yin2020-oq}.  

\par Magnetic interactions in frustrated kagome materials are of great complexity and result in complicated  noncollinear spin texture due to a delicate interplay between the symmetric Heisenberg exchange and asymmetric Dzyaloshinskii–Moriya interaction (DMI) which generates the spin canting.  Recently,  the Mn-based kagome magnets have attracted great attention stimulating vast investigations in the field of kagome materials.  Mn$_3$X (X = Rh, Ir, Pt, Ga, Ge, Sn) consisting of layered kagome planes have a noncollinear antiferromagnetic (AFM) spin texture \cite{PhysRevB.95.075128,  Park2018-zf,  Takeuchi2021-mn,  go2022non}.  SOC induces a virtual tilting of the magnetic moments and chiral spin texture induces spin currents in the absence of an external magnetic field \cite{go2022non}. 

\par RMn$_6$Sn$_6$ is a sibling of Mn-based kagome family (R=rare earth element) where the R elements play an important role in predicting the magnetic ground state.  The spin moments of the localized rare-earth R-4f and itinerant transition metal Mn-3d orbitals are inclined antiparallel \cite{SINNEMA1984333,  BROOKS199195} in heavy rare earth kagome systems, which leads to a ferrimagnetic ground state in (Gd, Tb, Dy,  Ho,  Er)Mn$_6$Sn$_6$ compounds \cite{MALAMAN1999519,  CLATTERBUCK199978,  Yin2020-ei}.  The magnetic coupling between the light R and Mn moment is parallel in (Nd,  Sm)Mn$_6$Sn$_6$ leading to a collinear FM ordering above room temperature \cite{PhysRevB.103.235109}.  Whereas,  the isostructural (Sc, Y, Lu)Mn$_6$Sn$_6$ with nonmagnetic R atoms  shows a spin spiral (SS) ordering of Mn moment  \cite{PhysRevB.103.235109,   PhysRevB.103.094413,  PhysRevB.103.014416,  sciadv.abe2680,  rosenfeld2008double,  PhysRevB.101.100405,  PhysRevB.104.024413}.  Therefore,  exploring the real space spin texture in its ground state and their relationship with kagome lattice geometry merit further study in RMn$_6$Sn$_6$ family,  especially when R is a non-magnetic.

\par YMn$_6$Sn$_6$ has a commensurate collinear structure above the room temperature which switches to a incommensurate phase upon cooling \cite{PhysRevB.101.100405}.  The incommensurate phase is very much sensitive to temperature and magnetic field \cite{sciadv.abe2680}.  However, YMn$_6$Sn$_6$ has been a subject of many recent experimental studies \cite{PhysRevB.103.094413,  PhysRevB.103.014416,  sciadv.abe2680,  rosenfeld2008double,  PhysRevB.101.100405,  PhysRevB.103.014416,  sciadv.abe2680,  PhysRevB.104.024413}.  The intrinsic anomalous Hall effect is stemming from the Berry curvature field in Mn-based distorted kagome lattice \cite{PhysRevB.104.024413}.  Therefore,  the relation between the topological electronic properties and complex magnetic behaviour in YMn$_6$Sn$_6$  has motivated wide interest in strongly correlated physics and makes a proliferate field to investigate the interplay between electron and spin.

\par The kagome magnets are candidates for quantum materials, but the distorted kagome magnets were rarely studied.  Unlike other members of the kagome magnet family,  YMn$_6$Sn$_6$ consists of two types of segregated Mn kagome planes.  Our spin dynamics simulation reports the magnetic ground state as a combination of two incommensurate SSs originating from two different sub-lattices of Mn atoms which is enforced due to exchange frustration along $c$-axis.  We capture the ferromagnetic (FM) - SS crossover with AFM spin exchange in YMn$_6$Sn$_6$.  We employed an effective model Hamiltonian  to understand the crossover physics of two non-equivalent and the incommensurate SSs microscopically.  Dirac crossings appear in the magnon bands of YMn$_6$Sn$_6$ which open topological gap with inclusion of DMI.  Our study is further collaborated by finding the Chern numbers and Berry curvature which confirm the existence of topological magnon in incommensurate SS phase of YMn$_6$Sn$_6$.

\begin{figure} [ht] 
\centering
\includegraphics[width=0.5\textwidth,angle=0]{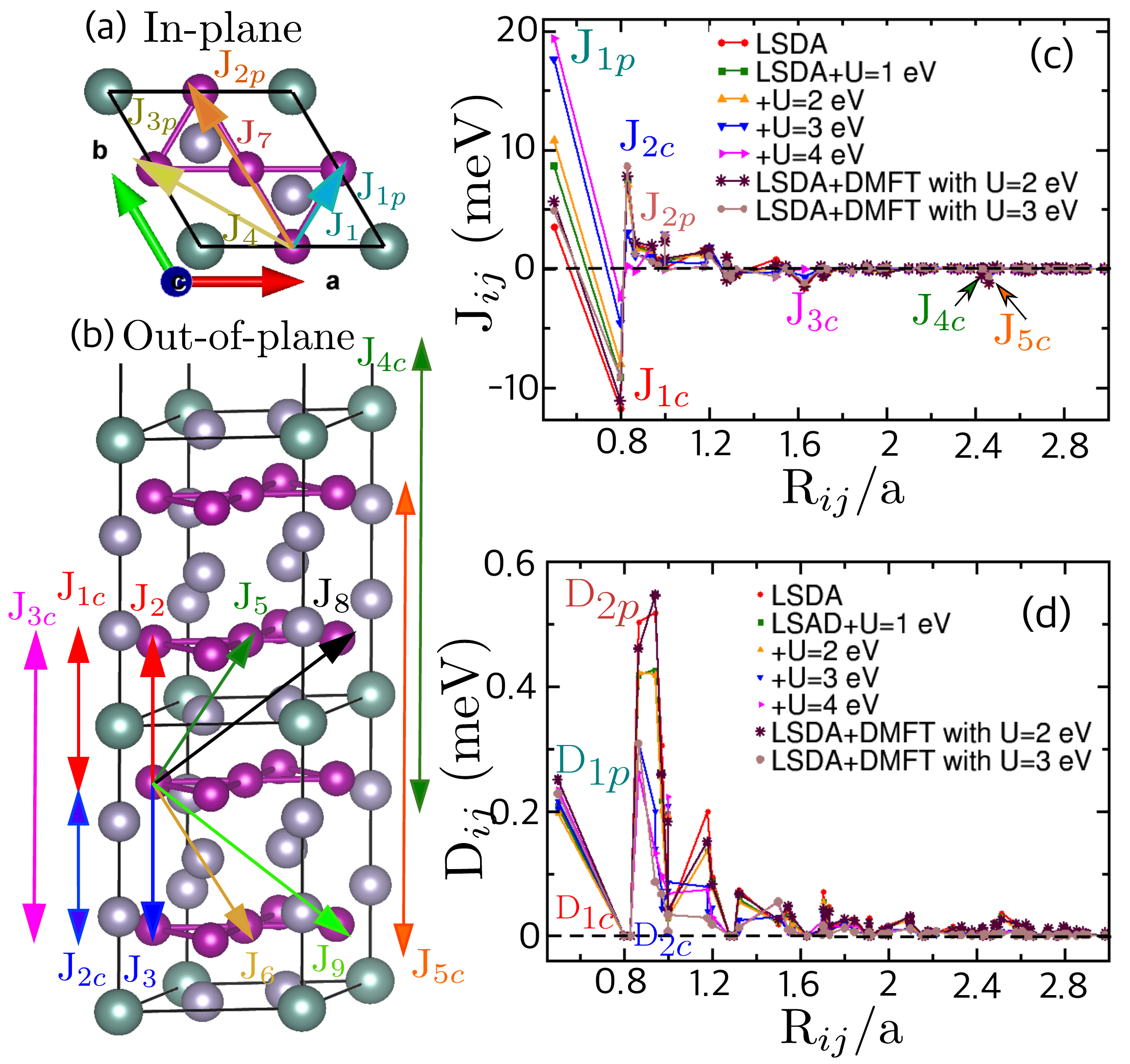} 
\caption{(a)-(b) Magnetic exchange interactions for nearest neighbour (NN) mapping in YMn$_6$Sn$_6$.  The atoms in purple, green and gray colours represent Mn, Y, Sn atoms respectively.  (c) Isotropic part of Heisenberg interactions ${\mathrm{J}}_{ij}$'s  and (d) anti-symmetric Dzyaloshinskii–Moriya interactions ${\mathrm{D}}_{ij}$'s in YMn$_6$Sn$_6$.  Here we used the convention of positive (negative) ${\mathrm{J}}_{ij}$'s as FM (AFM) and the subscript c (p) for $c$-axis ($ab$-plane) magnetic interactions throughout the text.  }
\label{fig1} 
\end{figure}

{\it{\rb {Complex magnetic interactions :}}} 
YMn$_6$Sn$_6$ consists of two inequivalent Mn based kagome planes stacking along the $c$-axis \cite{sm_text}. The two inequivalent planes are separated by Sn$_3$ and Sn$_2$Y layers. The Mn planes separated by Sn$_3$ are slightly larger in distance than Mn planes separated by Sn$_2$Y.  Because of this unique structural feature, the magnetic properties of YMn$_6$Sn$_6$  are very sensitive to both the distance between Mn-Mn along the $c$-axis and the nature of R element in the series of RMn$_6$Sn$_6$ (R=rare earth element).  The isotropic Heisenberg exchange interactions ${\mathrm{J}}_{ij}$'s have been calculated without and with SOC using different density functionals like spin polarized local density approximation (LSDA) and  general gradients approximation (GGA) which follow a long-ranged Ruderman-Kittel-Kasuya-Yosida (RKKY) coupling \cite{sm_text}.  The  full mapping of Mn-Mn magnetic exchanges in YMn$_6$Sn$_6$ for nearest neighbours are presented in the Fig.\ref{fig1}(a)-(b) along in-plane ($ab$-plane) and out of plane  ($c$-axis) directions respectively.  $ab$-plane Mn-Mn interaction (J$_1$) is strongly FM whereas the Mn-Mn interactions along the c-axis are markedly different following the AFM (${\mathrm{J}}_2$) coupling across the Sn$_2$Y layer and FM exchange interaction (${\mathrm{J}}_3$) across the Sn$_3$ layer.

\par To describe the exchange interactions in more efficient way,  we took  1$\times$1$\times$2 supercell of YMn$_6$Sn$_6$ as a smallest possible magnetic unit cell which contains 4 Mn kagome planes following AFM coupling across Sn$_2$Y layer and FM coupling across the Sn$_3$ layer.  The total magnetic exchange interactions in YMn$_6$Sn$_6$ are long ranged and can be categorized into two sets.  One is in the kagome plane ($ab$-plane) and other one is along the $c$-axis. The Heisenberg magnetic exchange interactions in the kagome plane ( denoted by J$_{ip}$ for i=1, 2, 3) are ${\mathrm{J}}_{1p}$ (${\mathrm{J}}_1$),  ${\mathrm{J}}_{2p}$(${\mathrm{J}}_{4}$),  ${\mathrm{J}}_{3p}$(${\mathrm{J}}_{7}$) and along the $c$-axis (denoted by ${\mathrm{J}}_{ic}$ for i =1, 2, 3, 4, 5) are ${\mathrm{J}}_{1c}$(${\mathrm{J}}_{2}$),  ${\mathrm{J}}_{2c}$(${\mathrm{J}}_{3}$),  ${\mathrm{J}}_{3c}$(${\mathrm{J}}_{16}$), ${\mathrm{J}}_{4c}$(${\mathrm{J}}_{n}$),  ${\mathrm{J}}_{5c}$(${\mathrm{J}}_{m}$) as shown in the Fig.\ref{fig1}(a)-(b) \cite{sm_text}.

\par Calculated magnetic exchange interactions for YMn$_6$Sn$_6$ are shown in the Fig. \ref{fig1}(c).  The in-plane Mn atoms have strong FM interaction (${\mathrm{J}}_{1p}= 3. 34 $ meV).  The isotropic Heisenberg exchange interactions along the $c$-axis (${\mathrm{J}}_{ic}$ for i=1, 2, 3, 4, 5) are -11.74,  8.66,  -1.61,  -0.37 and -1.23 meV respectively.   The inter and intra Mn-Mn  layers have AFM (J$_{1c}$) and FM (${\mathrm{J}}_{2c}$) coupling respectively,  whereas the intra and inter bilayers have AFM (${\mathrm{J}}_{3c}$) coupling.   The next inter (${\mathrm{J}}_{4c}$) and intra (J$_{5c}$) bilayers both have AFM coupling.  But the magnetic exchange interaction between Mn layers separated by Sn$_3$ layer are interacting FM (J$_{2c}$).  Therefore,  competition between the  ${\mathrm{J}}_{2c}$ (FM) and ${\mathrm{J}}_{4c}$ (AFM) leads to frustration in the structural geometry and a complex magnetic ground state in YMn$_6$Sn$_6$ \cite{sm_text}.  We will discuss in the subsequent section. The calculated magnetic moment is 2.36 $\mu_B$/Mn from LSDA.

\par Both the LDA and GGA fail to capture the magnetic exchange interactions accurately containing d-orbitals compared to experimental report \cite{PhysRevB.101.100405, sm_text} which requires the treatment of inclusion of Hubbard U on Mn-3d state.   Figure \ref{fig1}(c) shows the isotropic Heisenberg interactions J$_{ij}$'s with the inclusion of Hubbard U on Mn-3d orbital.   LDA+U affects both the in-plane ($ab$-plane) and out of plane (along $c$-axis) magnetic exchanges in YMn$_6$Sn$_6$.  ${\mathrm{J}}_{1p}$ increases with increasing the strength of Hubbard U on Mn-3d state.  The calculated in-plane (J$_{1p}$) and out of plane magnetic exchange interactions along the $c$-axis (${\mathrm{J}}_{ic}$ for i=1, 2, 3, 4, 5) are 10.59 meV,  and -8.01,  7.09,  -1.21,  -0.49, -1.07 meV respectively with Hubbard U = 2 eV.  The remarkable difference in the Heisenberg exchange interactions with inclusion of Hubbard U is that intra layer FM coupling (${\mathrm{J}}_{2c}$) decreases and next inter bilayer AFM coupling (${\mathrm{J}}_{4c}$) increases respectively which enhances frustration in the out of plane magnetic exchanges.  However,  ${\mathrm{J}}_{1p}$ and ${\mathrm{J}}_{1c}$,  ${\mathrm{J}}_{2c}$, ${\mathrm{J}}_{3c}$,  ${\mathrm{J}}_{4c}$,  ${\mathrm{J}}_{5c}$,  are 17.61 meV and  -4.51,  3.01  -0.66,  -0.33, -0.45  meV respectively from LSDA+U with a Hubbard U = 3 eV.  The calculated magnetic moment is 2.26 (2.21) $\mu_B$/Mn from LSDA+U with U = 2 (3) eV.

\par The correlation effects among the localised Mn-3d electrons play an important role in the kagome lattice. This motivate us to use a combination of DFT and dynamical mean-field theory (DMFT) with inclusion of Hubbard U on Mn-3d state to describe the electronic structure properly.  We used the LSDA functional and spin-polarized T-matrix fluctuation impurity solver to accurately capture the magnetic exchanges as shown in Fig.\ref{fig1}(c) (for U = 2 and 3 eV respectively).  ${\mathrm{J}}_{1p}$ and ${\mathrm{J}}_{1c}$,  ${\mathrm{J}}_{2c}$, ${\mathrm{J}}_{3c}$,  ${\mathrm{J}}_{4c}$,  ${\mathrm{J}}_{5c}$,  are 5.49 meV and  -11.13,  7.81,  -1.49,  -0.69,  -1.22 meV respectively from LSDA+DMFT with a Hubbard U = 2 eV.  However,  ${\mathrm{J}}_{1p}$ and ${\mathrm{J}}_{1c}$,  ${\mathrm{J}}_{2c}$, ${\mathrm{J}}_{3c}$,  ${\mathrm{J}}_{4c}$,  ${\mathrm{J}}_{5c}$,  are 4.85 meV and  -9.17,  8.61,  -1.18,  -0.02, -0.65  meV respectively from LSDA+DMFT with a Hubbard U = 3 eV.  How ${\mathrm{J}}_{ic}$'s affect the spin texture for the magnetic ground state of  YMn$_6$Sn$_6$ that will be discussed in the subsequent section. The calculated magnetic moment is 2.29 (2.23) $\mu_B$/Mn from LSDA+DMFT with U = 2 (3) eV is in excellent agreement with experimental report \cite{PhysRevB.101.100405}.

\par Symmetric Heisenberg exchanges ${\mathrm{J}}_{ij}$'s indulge collinear ordering whereas the anti-symmetric DMI's (${\mathrm{D}}_{ij}$'s) cant the spin favouring the non-collinear magnetic ground state.  Figure \ref{fig1}(d) show DMI for YMn$_6$Sn$_6$.  ${\mathrm{D}}_{1}$(${\mathrm{D}}_{1p}$) increases whereas ${\mathrm{D}}_{4}$(${\mathrm{D}}_{2p}$) has a continuous decreasing trends with increasing Hubbard U.  The inter (${\mathrm{D}}_{2}$ = ${\mathrm{D}}_{1c}$) and intra layer (${\mathrm{D}}_{3}$ = ${\mathrm{D}}_{2c}$) DMIs are zero.  The components DMIs along the in-plane ($ab$-plane) and out of plane ($c$-axis) directions for ${\mathrm{D}}_{1p}$ are ${{\mathrm{D}}_{1p}}^{ \parallel}$ $\sim 0. 251 (0.218) $ meV and ${{\mathrm{D}}_{1p}}^z$ $\sim 0$ meV respectively from LSDA+DMFT with U = 2 (3) eV.  Whereas they are ${{\mathrm{D}}_{2p}}^{ \parallel}$ $\sim 0. 068 (0.046)$  meV and ${{\mathrm{D}}_{2p}}^z$ $\sim 0.455 (0.308)$ meV for ${\mathrm{D}}_{2p}$ respectively as shown in Fig.\ref{fig1}(d).


{\it{\rb {Frustration driven spin spiral :}}}  Capturing spin texture in the magnetic ground state of a  frustrated 3D kagome lattice like YMn$_6$Sn$_6$ is a real challenge and needs assiduous attention.  The system consists of stacking of FM kagome planes where the frustration in magnetic exchanges come from the next inter bilayer couplings (J$_{4c}$).  Real-space magnetic configurations are confirmed by both Monte carlo (MC) simulations and spin dynamics (SD) simulation as implemented in UppASD code \cite{uppasd} with the spin Hamiltonian given by :
\begin{equation}
    {\mathrm{H}}= -\sum_{i,j} {\mathrm{J}}_{ij} {\bm {\mathrm{S}}}_i \cdot {\bm {\mathrm{S}}}_j - \sum_{i,j} {\bm {\mathrm{D}}}_{ij} \cdot ({\bm {\mathrm{S}}}_i \times {\bm {\mathrm{S}}}_j) -\sum_i {\mathrm{K}}({{\mathrm{S}_{i}^{z}}})^2,\nonumber
\end{equation}
Where ${\mathrm{K}}$ is the magnetic anisotropic energy. The calculated magnetic anisotropy energy (E$_c$-E$_p$) for YMn$_6$Sn$_6$ is $\sim 0.05$ meV from LSDA+DMFT with Hubbard U = 2 eV which is below to the experimental report ($\sim 0.2$ meV) \cite{sciadv.abe2680}.  

\par We simulate a finite-sized 3D kagome slab having dimension with  96$\times$96$\times$31 with periodic boundary conditions.   Efficient thermalization is guaranteed by simulated temperature annealing where the both the MC and SD simulations are started from a random spin configuration corresponding to high temperature 1000 K until the lowest temperature 10$^{-5}$ K is reached.  Raising the temperature randomizes spin orientation due to thermal fluctuations to avoid any specific magnetic order.   At each temperature,  we used 2$\times$10$^{5}$ MC sweeps for equilibration, and 5$\times$10$^{5}$ MC sweeps (in time steps of steps of 1000 MC sweeps and in time steps 10$^{-16}$ sec) for the measurement of physical observable.  In order to search a correct magnetic ground state,  metastable states at the low-temperature regime are avoided by starting the simulation from a variational ground state and then increasing the temperature using the MC scheme.  We used the magnetic exchanges and DMI parameters for the spin Hamiltonian from three different calculation schemes like LSDA,  LSDA with Hubbard U = 2,  3 eV and DFT+DMFT with Hubbard U = 2,  3 eV respectively.

\begin{figure} [ht] 
\includegraphics[width=0.5\textwidth,angle=0]{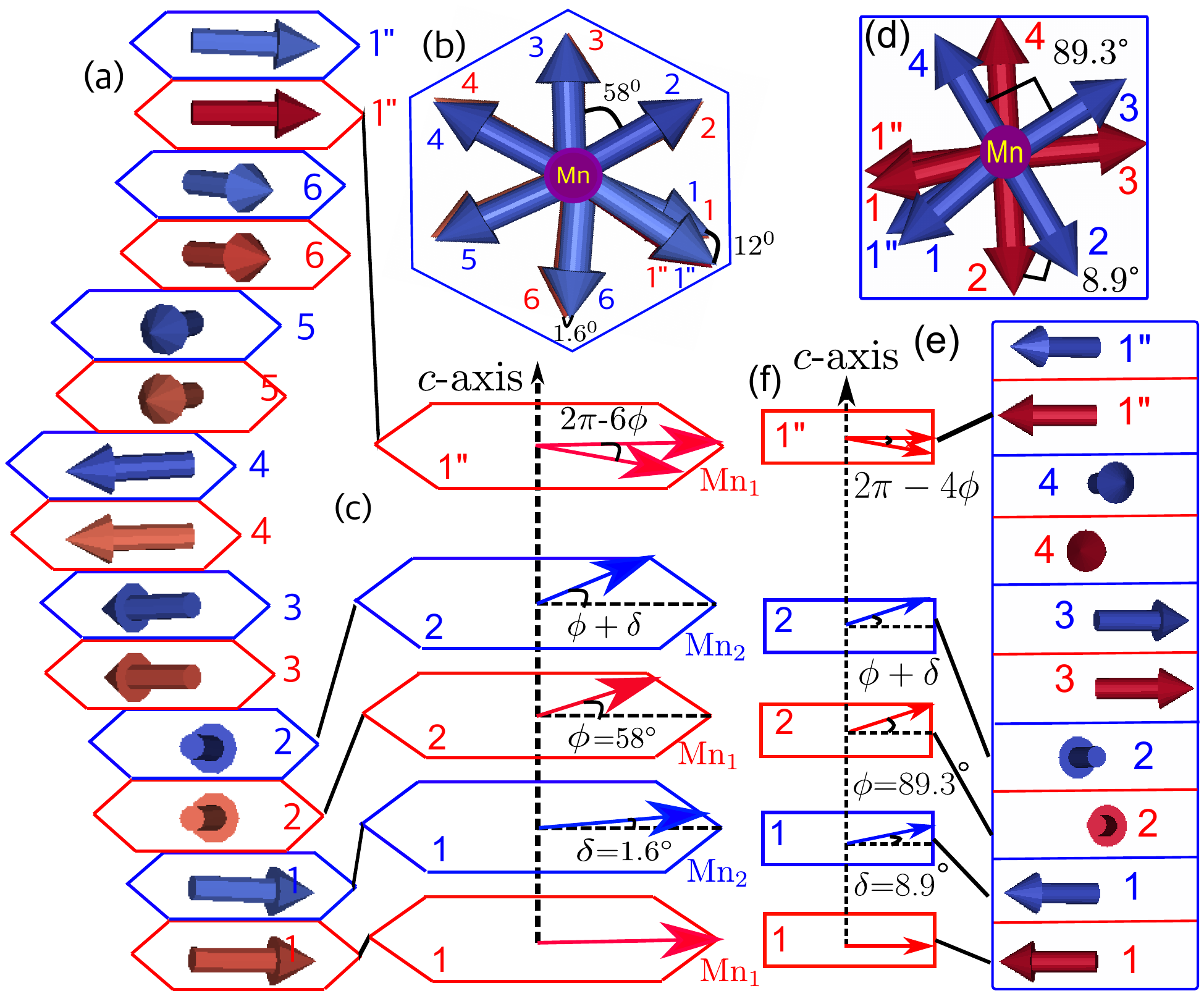} 
\caption{(a) Side and (b) top view of two incommensurate spin spirals (SS) propagating along $c$-axis marked by two different colors from spin dynamics simulation with magnetic interactions from LSDA+DMFT with U = 2 eV.  The SS repeats after 6.19 Mn layers marked by numbers. (c) Schematic representation of two SSs with an angle $\delta \sim 1.6^{\circ}$ between them and showing the pitch angle ($\phi \sim$58$^{\circ}$), and incommensurate angle (2$\pi-6\phi \sim$12$^{\circ}$) .  (d) Side, (e) top view and (f) schematic representation of  the same SSs with magnetic interactions from LSDA+DMFT with U = 3 eV.  Here the pitch angle $\phi$ and  $\delta$ are $89.3^{\circ}$ and $8.9^{\circ}$ respectively. The incommensurate angle reduces to 2$\pi-4\phi$ ($\sim$3$^{\circ}$) repeating after 4.03 Mn layers. 
}
\label{fig2} 
\end{figure}

\par Figure \ref{fig2}(a)-(b) shows the spin texture of magnetic ground state of YMn$_6$Sn$_6$ from LSDA+DMFT with U = 2 eV within full relativistic limit.  The magnetic ground state of YMn$_6$Sn$_6$  is a combination of two incommensurate SSs along the $c$-axis with a wave vector $\bf{q}$ $\sim$ (0 0 0.1615) and has a rotational plane along [$001$]. The incommensurate  spiral phase has been reported in separate experimental studies \cite{PhysRevB.101.100405,  sciadv.abe2680, rosenfeld2008double, PhysRevB.103.014416, PhysRevB.103.094413}, 
but demands the theoretical inspection from a combination of full first principles and spin dynamics approach for the prediction of correct magnetic ground state.  The two incommensurate spirals (marked by blue and red colours in the Fig.\ref{fig2}(a)-(b)) correspond to two different Mn kagome planes as described in the magnetic crystal structure of YMn$_6$Sn$_6$ \cite{sm_text}.  The two adjacent moment layers belonging to two different Mn kagome planes are coupled FM with an angle $\delta$ $\sim$1.6$^{\circ}$ which is basically the angle between two incommensurate SSs as shown in Fig.\ref{fig2}(b)-(c).

\par Now we analyse the spin texture layer by layer.  Figure \ref{fig2}(a) and (b) show layer and top view spin texture of two incommensurate SSs respectively.  The two incommensurate SSs form by a combination of two cycloidal SSs (also called double fan like structure \cite{PhysRevB.96.104405}). The spin arrangements in each incommensurate SS repeats after 6.19 Mn layers along the $c$-axis marked by numbering in two different colours as shown in the Fig.\ref{fig2}(a)-(b) with a small incommensurate angle.  The pitch angle ($\phi$) and incommensurate angle (2$\pi-6\phi$)  in each cycloidal SS or fan moment in this double-fan like structure are $\sim$58$^{\circ}$ and $\sim$12$^{\circ}$ respectively.

\par The SSs in YMn$_6$Sn$_6$ are sensitive on the choice oh Hubbard U on Mn-3d orbitals.   The pitch angle and $\delta$ of two incommensurate SSs increase to $\sim$89.3$^{\circ}$ and $\sim$8.9$^{\circ}$ if we increase the Hubbard U on Mn-3d orbitals to 3 eV in LSDA+DMFT as shown in Fig.\ref{fig2}(d)-(f).  The spin arrangements in each incommensurate SS repeats after 4.03 Mn layers with LSDA+DMFT with a Hubbard U = 3 eV.  Here the incommensurate angle (2$\pi-4\phi$) and wave vector ($\bf{q}$) for each cycloidal SS are $\sim $ 3$^{\circ}$ and $\sim$ (0 0 0.248) respectively which is good agreement with the experimental reports \cite{Ghimire2020-qs, PhysRevB.104.024413, rosenfeld2008double,  PhysRevB.101.100405}.

\par We also studied the effects of anisotropic energy on the incommensurate SS within relativistic limit.  The anisotropic energy flips the rotational plane of double-fan structure from [$1\bar{1}0$] to [$001$] and locks it in $ab$-plane. The incommensurate SS is not DMI induced, rather it is inter bilayer frustration enforced.  The effect of exchange frustration is less in magnetic interactions calculated from LSDA. Therefore,  LSDA gives a FM ground state consisting of AFM block with 4 Mn layers along the $c$-axis where Mn kagome planes separated by  Sn$_2$Y and Sn$_3$ have AFM and FM spin arrangements respectively \cite{sm_text}.  Inclusion of a small Hubbard U $\sim$ (2-3 eV has a tendency of forming SS along the $c$-axis \cite{sm_text}.

\par The  root of the spin reorientation will be discussed in the next section within the framework of nearest neighbour magnetic exchange interactions along the $c$-axis (${\mathrm{J}}_{ic}$ for i =2, 4).   YMn$_6$Sn$_6$ is three dimensional (3D) kagome metal of having long range RKKY interactions where magnetic exchanges along $c$-axis play the main role of predicting the complex magnetic ground state.   The in-plane magnetic exchanges in kamoge plane are always interacting ferromagnetically.  The inter and next intra bilayer magnetic exchanges in YMn$_6$Sn$_6$ are ${\mathrm{J}}_{2c}$ $\sim$ 8.66 meV and ${\mathrm{J}}_{4c}$ $\sim$ -0.37 meV respectively from LSDA.   Whereas the corresponding values are ${\mathrm{J}}_{2c}$ $\sim$ 8.61 (7.81) meV and  ${\mathrm{J}}_{4c}$ $\sim$ -0.02 (-0.69) meV from LSDA+DMFT with a Hubbard U = 3 (2) eV.  The competition between ${\mathrm{J}}_{2c}$ and  ${\mathrm{J}}_{4c}$ tends to form two SSs with a plane of rotation along [$001$] \cite{sm_text}.  It gives two incommensurate SSs with a pitch angle $\phi$  $\sim$ 89.3$^{\circ}$ (58$^{\circ}$) and $\delta$ $\sim$ 8.9$^{\circ}$ (1.6$^{\circ}$) with LSDA+DMFT with a Hubbard U = 3 (2) eV where anisotropy energy fastens the SS strongly in the kagome plane of YMn$_6$Sn$_6$ as shown in Fig.\ref{fig2}.  Putting ${\mathrm{J}}_{4c}$ $\sim$ 0, the pitch angle $\phi$ remains same but the angle between two incommensurate SSs $\delta$ increases to $\sim$  11$^{\circ}$ ($\pm$0.2).  The out of plane magnetic exchanges for YMn$_6$Sn$_6$ are ${\mathrm{J}}_{2c}$ $\sim$ 3.01 (7.09) meV and  ${\mathrm{J}}_{4c}$ $\sim$ -0.33 (-0.49) meV from LSDA+U with U = 3 (2) eV.  However,  the pitch angle $\phi$ of the incommensurate SS is 102$^{\circ}$ ($\sim$54$^{\circ}$) from LSDA+U with U = 3 (2) eV respectively.

{\it{\rb {Effective Hamiltonian for incommensurate SSs :}}} 
Spin dynamics simulation fails to capture the experimentally observed two non-equivalent SS with slightly different pitch angles \cite{sciadv.abe2680}.  Therefore, we consider an effective model Hamiltonian for the two incommensurate SSs considering total the magnetic exchange interactions (called it an effective exchanges) within Mn planes along $c$-axis (J$_{ic}^{eff}$ with i =1,2,3,4,5)  \cite{sm_text} to study the effect of Hubbard U on SSs of YMn$_6$Sn$_6$ and is given by :
\begin{eqnarray}
{\mathrm{H_{ic}}^{eff}}= - {\mathrm{J}}_{1c}^{eff} {\cos(\phi - \delta)} - {\mathrm{J}}_{2c}^{eff} {\cos\delta} - 2{\mathrm{J}}_{3c}^{eff} {\cos\phi} \nonumber \\ - {\mathrm{J}}_{4c}^{eff} {\cos(2\phi - \delta)} - {\mathrm{J}}_{5c}^{eff} {\cos(\phi + \delta)} \nonumber 
\label{effc-ham}
\end{eqnarray}
\begin{figure} [ht] 
\includegraphics[width=0.5\textwidth,angle=0]{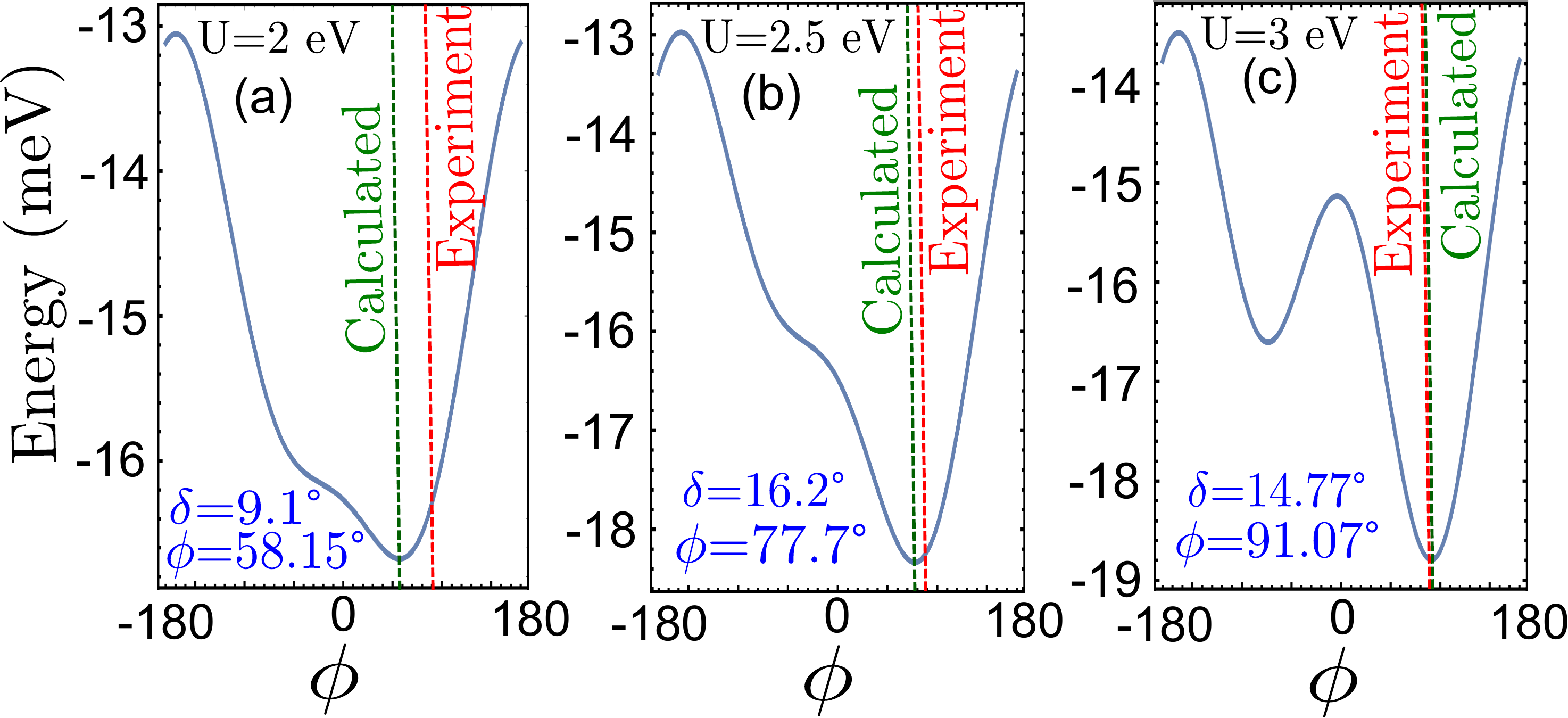} 
\caption{Calculated variation of energy with the pitch angle from LSDA+DMFT with (a) U = 2 eV,  (b) U = 2.5 eV and (c) U = 3 eV respectively. The calculated pitch angles are compared with experimental report \cite{sciadv.abe2680}.
}
\label{fig3} 
\end{figure}
Such qualitative considerations allow us to understand the detail root of arising SS in the YMn$_6$Sn$_6$ and evolution of pitch angles of two incommensurate SSs with correlation. The effective magnetic interactions within Mn planes along the $c$-axis  are interacting FM which leads to a FM ground state from LSDA calculations ($\phi=$0, $\delta$=0) \cite{sm_text}.  The inclusion of Hubbard U on Mn-3d orbitals leads to SS magnetic ground state from FM for both LSDA+U and LSDA+DMFT.  We scan the magnetic ground state of YMn$_6$Sn$_6$ with different Hubbard U parameters ranging from U = (1-4) eV  (see table II in supplementary material) \cite{sm_text}.  YMn$_6$Sn$_6$ remains in spiral phase with U = (2-3) eV and remains in FM phase otherwise.  Figure \ref{fig3}(a)-(c) represent the energy evolution of SS with pitch angle with Hubbard U = 2, 2.5, 3 eV respectively.  The asymmetric double well nature of energy with pitch angles confirms the signature of two non-equivalent and incommensurate SSs in YMn$_6$Sn$_6$.  The calculated pitch angle for the dominant SS $\phi$  $\sim$ 91$^{\circ}$  matches with the experimental reports \cite{Ghimire2020-qs, PhysRevB.104.024413, rosenfeld2008double,  PhysRevB.101.100405} which is also in good agreement with the simulated SS ($\phi$  $\sim$ 89$^{\circ}$  ) obtained from LSDA+DMFT with U = 3 eV.

\par In order to understand the effect of correlation on incommensurate SS,  we analyse the effective magnetic exchanges (J$_{ic}^{eff}$'s) with different Hubbard U.  J$_{1c}^{eff}$, J$_{2c}^{eff}$ are  FM whereas J$_{5c}^{eff}$ is AFM always.  Now exchange frustration in YMn$_6$Sn$_6$ comes from  J$_{3c}^{eff}$, J$_{4c}^{eff}$ when they switches into AFM from FM with gradually increasing Hubbard U on Mn-3d orbital (see table II in supplementary material) \cite{sm_text}.  The inclusion of Hubbard U = 2 eV  flips J$_{4c}^{eff}$ to AFM (-0.54 meV) from FM but J$_{4c}^{eff}$ remains FM (0.13 meV).  This leads to frustration in the magnetic exchanges of YMn$_6$Sn$_6$ and gives magnetic ground state a combination of two incommensurate SSs with $\phi$ = 58.15$^{\circ}$, $\delta$ = 9.1$^{\circ}$.   The exchange frustration increases with U=2.5 eV and J$_{3c}^{eff}$ flips into AFM (-0.34 meV) from FM which increase both the pitch angle $\phi$ and $\delta$ to  77.7$^{\circ}$ and 16.2$^{\circ}$ respectively.  YMn$_6$Sn$_6$ gives SS with $\phi$ = 91.07$^{\circ}$ and $\delta$ = 14.77$^{\circ}$ in presence of U = 3 eV where J$_{3c}^{eff}$ (-1.32 meV) and J$_{4c}^{eff}$ (-1.62 meV) are strongly AFM as shown in Fig.  \ref{fig3}(c).


{\it{\rb {Topology in magnon:}}} The kagome lattice provides an ideal system to investigate topological magnon.  AFM kagome lattice exhibits a flat spin wave mode due to geometrical spin frustration \cite{PhysRevLett.96.247201} whereas a flat mode also exists in a FM kagome \cite{PhysRevLett.115.147201}. However,  the topological investigation in kagome SS has pending yet.  We calculated the magnon bands along the high symmetry $\Gamma-{\mathrm{K}}-{\mathrm{M}}-\Gamma$ direction from LSDA+DMFT without and with SOC.  Two Dirac crossings (Dirac magnons) appear at 85 and 108 meV (with U = 3 eV) at high symmetry ${\mathrm{K}}$ point without SOC where the bands are doubly degenerated.  However, the two Dirac points appear slightly below at 73 and 100 meV with U = 2 eV \cite{sm_text}.  To compare with calculated  magnon spectra with another experimental study reported by  Zhang  et al. \cite{PhysRevB.101.100405},  we calculated the dynamical structure factor ${\mathrm{S}}({\mathrm{q}},\omega)$ from spin dynamics simulation at T = 5.  ${\mathrm{S}}({\mathrm{q}},\omega)$ is proportional to the inelastic scattering intensity which is in good agreement with the experimental measured spectra by Zhang  et al. \cite{PhysRevB.101.100405} with LSDA+DMFT with U = 2 eV \cite{sm_text}.

\begin{figure} [ht] 
\includegraphics[width=0.5\textwidth,angle=0]{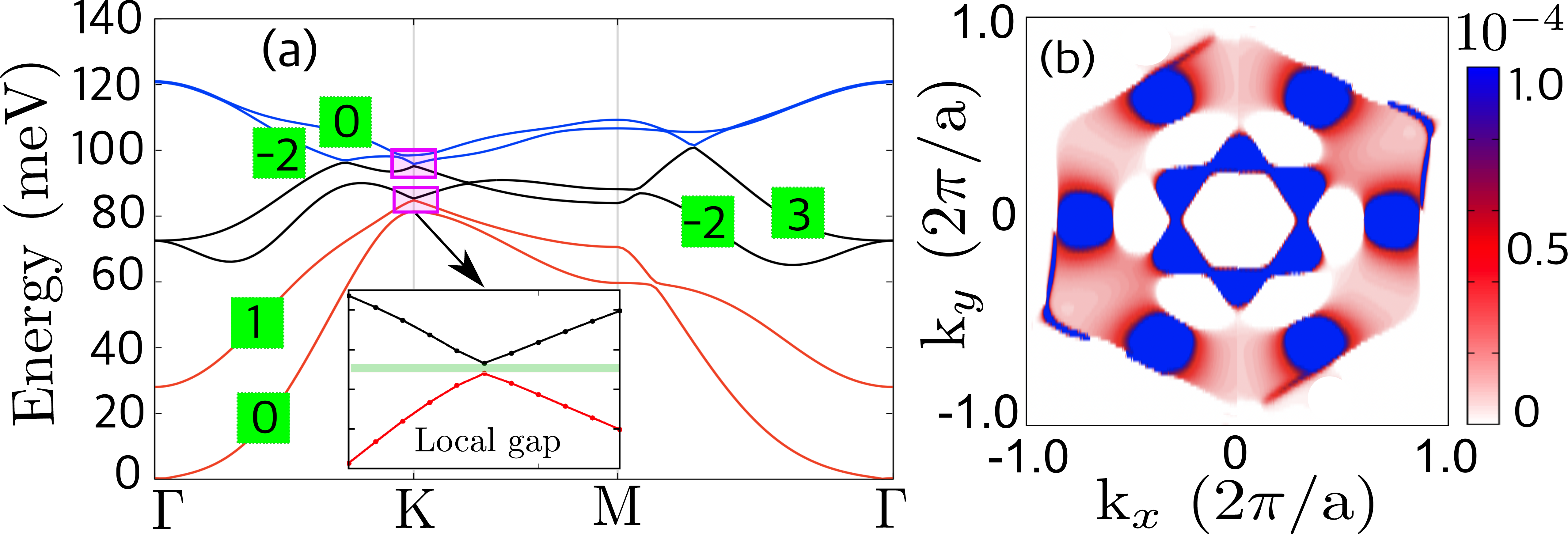} 
\caption{(a) Magnon bands with calculated Chern numbers and (b) Berry curvature profile in YMn$_6$Sn$_6$ with SOC. Topological gaps are marked by boxes in figure (a).}
\label{fig4} 
\end{figure}

\par Figure \ref{fig4}(a) shows the magnon bands of YMn$_6$Sn$_6$ where the magnetic ground state is a admixture of two incommensurate SSs.  We can categorized the total bands into three sets (upper, middle and lower bands) coming from three distinct magnon bands in kagome lattice.  The degeneracies at ${\mathrm{K}}$ point (Dirac crossings) have broken with inclusion of DMI (with SOC) and two topological gap (topological magnon) appear at that point.  One gap is of 0.52 meV around 84 meV and other is of 0.64 meV around 95 meV (marked by boxes in Fig.\ref{fig4}(a)).  Similar to magnon,  we also observed Dirac points which open gap in presence of SOC and flat bands near the Fermi level in the electronic band structure of YMn$_6$Sn$_6$ \cite{sm_text}.

\par Our findings are further corroborated by calculating the topological invariants for acid testing of the existence of topological magnon in YMn$_6$Sn$_6$.  There exist no global gap in magnon bands as the maximum of the each band is higher in energy than the minimum of its next magnon band. Therefore, one can define the Chern numbers considering local gaps around the high symmetry ${\mathrm{K}}$ point in Brillouin zone (BZ) where topological gap appear \cite{PhysRevLett.122.187203, PhysRevLett.106.016402,  PhysRevB.106.125112}. The calculated Chern numbers are 0, 1,  -2,  3,  -2 0 respectively as shown in the Fig.\ref{fig4}(a).  We got the maximum Chern number  $ c = -2$ for upper magnon band belonging to flat band of kagome lattice as the number of Mn kagome planes in an unit cell is 2 \cite{PhysRevB.86.241111}.  Figure \ref{fig4}(b) shows the calculated Berry curvature profile. Here the maxima appears at high symmetry ${\mathrm{K}}$ point where the topological gap appears in BZ. This is another signature of existing topological magnon in kagome SS of YMn$_6$Sn$_6$.

\par We further investigate the effects of DMI's on the existence of topological magnon considering an quasi 2D structure of YMn$_6$Sn$_6$.  Here consider only in-plane ($ab$-plane) magnetic interactions (${\mathrm{J}}_{ip}$,  ${\mathrm{D}}_{ip}$ for i=1, 2 as both ${\mathrm{D}}_{1c }$ and  ${\mathrm{D}}_{2c }$ are 0) on topological magnons.  The topological insulating phase in kagome magnet switches into gapless topological mode for quasi 2D structure ($\frac{{{\mathrm{D}}_{1p}}^z}{{\mathrm{J}}_{1p}} \sim 0$) \cite{sm_text} which opens a gap with inclusion of ${{\mathrm{D}}_{2p}}^z$ $\neq$.  Therefore,  opening of two topological gaps shown in Fig. \ref{fig4}(a) are mainly attributed by the ${{\mathrm{D}}_{2p}}^z$ $\neq$ 0 as ${{\mathrm{D}}_{1p}}^z \sim 0$.  However, $ab$-plane magnetic interactions (${\mathrm{J}}_{ip}$,  ${\mathrm{D}}_{ip}$) within a kagome plane play the governing role in producing the topological magnon whereas the magnetic exchanges along the $c$-axis (${\mathrm{J}}_{ic}$) introduce incommensurate SS phase in YMn$_6$Sn$_6$.


{\it{\rb {Conclusion and outlook :}}} We demonstrate the reorientation of spin texture from the FM state into two incommensurate SSs in YMn$_6$Sn$_6$ due to next inter bilayer magnetic exchange frustration.  YMn$_6$Sn$_6$ has a FM ground state from DFT approach,  whereas has a SS phase from DFT+DMFT+U approach.  The reason for this discrepancy in predicting experimental ground stems from the failure of DFT to capture the effect of electronic correlations.  We propose a model Hamiltonian for two incommensurate SS to explain FM-SS crossover due to AFM spin exchange with correlation.  We reported the topological magnon in the kagome SS of YMn$_6$Sn$_6$ emerging from in-plane magnetic exchanges.   Additionally,  we observed DPs and flat bands near the Fermi level in the electronic structure of YMn$_6$Sn$_6$.  Our study provides a detailed platform to understand the influence of the out of plane Heisenberg exchange interactions on the behaviour of  incommensurate SSs,  and DMIs stemming from the Mn-based kagome planes on topological magnon respectively.  The interplay between magnetic structure and topology in both fermions (spin) and bosons (magnon) makes YMn$_6$Sn$_6$ as an ideal candidate to study the topological Hall transport in future \cite{Liu2018, PhysRevB.103.235109,  PhysRevB.104.024413,  PhysRevB.107.L081110, PhysRevB.89.134409}.

\begin{acknowledgments}
BS would like to thank H.  Zhang for providing the experimental cif file,  Y. O. Kvashnin and I. Miranda for fruitful discussions.
Financial support from Vetenskapsrådet (grant numbers VR 2016-05980 and VR 2019-05304), and the Knut and Alice Wallenberg foundation (grant numbers KAW 2018.0060, KAW 2021.0246, and KAW 2022.0108) is acknowledged.
A.B. acknowledges eSSENCE.  
The computations were enabled by resources provided by the National Academic Infrastructure for Supercomputing in Sweden (NAISS) and the Swedish National Infrastructure for Computing (SNIC) at NSC and PDC, partially funded by the Swedish Research Council through grant agreements no. 2022-06725 and no. 2018-05973.
\end{acknowledgments}

\bibliography{yms}{}


\clearpage
\begin{onecolumngrid}
	\begin{center}
		{\fontsize{12}{12}\selectfont
			\textbf{Supplementary material for ``Topological magnon in exchange frustration driven incommensurate spin spiral of a kagome lattice YMn$_6$Sn$_6$"\\[5mm]}
		{\normalsize Banasree Sadhukhan,$^{1,2}$,  Anders Bergman,$^{3}$,  Patrik Thunstr{\"o}m,$^{3}$,  Manuel Pereiro Lopez $^{3}$,  Olle Eriksson,$^{3}$,  Anna Delin, $^{2, 4,5}$\\[1mm]}
		{\small $^1$\textit{Department of Physics and Nanotechnology, SRM Institute of Science and Technology, Kattankulathur, 603203, Chennai, Tamil Nadu, India}\\[0.5mm]}
		{\small $^2$\textit{Department of Applied Physics, School of Engineering Sciences, KTH Royal Institute of Technology, AlbaNova University Center, SE-10691 Stockholm, Sweden}\\[0.5mm]}
		{\small $^3$\textit{Department of Physics and Astronomy, Uppsala University, Box 516, SE-75120 Uppsala, Sweden}\\[0.5mm]}
		{\small $^4$\textit{Swedish e-Science Research Center (SeRC), KTH Royal Institute of Technology, SE-10044 Stockholm, Sweden}\\[0.5mm]}
		{\small $^5$\textit{Wallenberg Initiative Materials Science for Sustainability (WISE), KTH Royal Institute of Technology, SE-10044 Stockholm, Sweden}\\[0.5mm]}
		}
	\end{center}
	
\end{onecolumngrid}


\section{Theoretical and computational details}

\subsection{Magnetic interactions}

\par The density functional theoretical (DFT) calculations for both the electronic structures and the magnetic interactions are performed within full-potential linear muffin-tin orbital-based code RSPt \cite{wills2010full} using both local spin density approximation (LDA) and general gradient approximation (GGA) functionals.  We used 18$\times$18$\times$18 kmesh for all DFT calculations.  We further used a combination of DFT (spin polarized version of LDA)  and dynamical mean-field theory (DMFT) with spin-polarized T-matrix fluctuation impurity solver as implemented in RSPt code  to correctly capture the correlation of Mn-$3d$ orbitals for describing electronic structure.  We used a k-mesh of 12$\times$12$\times$12 for LSDA+DMFT to reduce the computational cost. The convergence with k-mesh used for integrations in BZ has been carefully checked for both DFT and DMFT.

\par The Heisenberg Hamiltonian describing the magnetic system is given by 
\bea
 \mathrm{H} = -\sum_{ij} \sum_{\{\alpha,\beta\}} e^{\alpha}_i \mathrm{J}^{\alpha \beta}_{ij} e^{\beta}_j 
\label{eq:HamS}
\eea
where $e^{\alpha}_i$ ($e^{\beta}_j$) is the $\alpha$ ($\beta$) component of the unitary vector pointing along the direction of the spin located at the site $i$ ($j$).  Considering   $\mathrm{J}^{\alpha \beta}_{ij}$ as a [3$\times$3] matrix, the isotropic (Heisenberg) part of the magnetic exchange interactions J$_{ij}$'s

and anti-symmetric Dzyaloshinskii–Moriya interactions (DMIs) D$_{ij}$'s are defined by 
\begin{eqnarray}
\mathrm{J}_{ij}= ({\mathrm{J}_{ij}}^{xx}+{\mathrm{J}_{ij}}^{yy}+{\mathrm{J}_{ij}}^{zz})/3,  \nonumber
\\ \nonumber
\mathrm{D}_{ij} = \mid \vec D_{ij} \mid = \sqrt{({\mathrm{D}_{ij}}^x)^2 + ({_{ij}}^y)^2 + ({D_{ij}}^z)^2}, \\
\label{def-int}
\end{eqnarray}
where $\mid \vec D \mid $ is the magnitude of the DMI vector.  Here we used the convention of positive J$_{ij}$'s as ferromagnetic (FM) and negative J$_{ij}$'s as antiferromagnetic (AFM).

\par For a given real material, the parameters in expression~\eqref{def-int} can be extracted from magnetic force theorem based on linear-response theory.  This is originally formulated for the case of isotropic Heisenberg interactions in the absence of spin-orbit coupling (SOC) \cite{liechtenstein1987local, PhysRevB.61.8906}.   The theory is  formulated for second order perturbation in the deviations of spins from equilibrium magnetic configuration.  The approach has been extended to take into account relativistic effects to allow to compute the full interaction tensor for $\mathrm{J}^{\alpha \beta}_{ij}$  \cite{antropov1997exchange,  PhysRevB.68.104436,  PhysRevB.79.045209,  secchi2015magnetic}.

\par Here we present a derivation of the formulae based on Green's functions formalism below.  We begin by perturbing the spin system by deviating the initial moments ($\vec e_0$) on a small angle $\vec{\delta \varphi}$ (the site index is omitted at the moment):
\bea
\vec e = \vec e_0 + \delta \vec e + \delta^2 \vec {e} = \vec e_0 + \bigl[ \vec{\delta \varphi} \times \vec e_0 \bigl] -\frac{1}{2} \vec e_0 (\vec {\delta \varphi})^2 \nonumber
\eea
Then one can write the Hamiltonian Eq.~\eqref{eq:HamS} of the perturbed system in terms of series in the order of $\vec \delta \varphi$:
\bea
\hat H' = \hat H^{(0)} + \hat H^{(1)} + \hat H^{(2)}. \nonumber
\eea
In the collinear limit,  all spins point along the same direction, which we set parallel to $Z$ axis. Then the tilting vectors have the following components:
\bea
\vec{\delta\varphi} = (\delta\varphi^x ; \delta\varphi^y ; 0)  \nonumber \\
\bigl[ \vec{\delta\varphi} \times \vec e_0 \bigl] = (\delta \varphi^y ; - \delta\varphi^x ; 0) \nonumber
\eea
Focusing on the energy contributions of the second order in $\vec{\delta \varphi}$ (i.e. $\hat H^{(2)}$), we obtain:
\bea
\mathrm{H}^{(2)} = -\sum_{i \ne j} \biggl( \mathrm{J}^{xx}_{ij} \delta\varphi_i^y \delta\varphi_j^y + \mathrm{J}^{yy}_{ij} \delta\varphi_i^x \delta\varphi_j^x - \mathrm{J}^{xy}_{ij} \delta\varphi_i^y \delta\varphi_j^x \nonumber \\ 
- \mathrm{J}^{yx}_{ij} \delta\varphi_i^x \delta\varphi_j^y -\frac12 \mathrm{J}^{zz}_{ij} ((\vec{\delta \varphi_i})^2 + (\vec{\delta \varphi_j})^2) \biggl) \nonumber
\eea
Then one can do the same perturbation for the electronic Hamiltonian ($\mathcal{H}$), which will become:
\bea
\hat{\mathcal{H}'} = \hat U^\dagger  \hat{\mathcal{H}} \hat U = \hat{\mathcal{H}}^{(0)} + \hat{\mathcal{H}}^{(1)} + \hat{\mathcal{H}}^{(2)}, \nonumber
\eea
where $\hat U=\exp{(i\vec{\delta\varphi} \hat{\vec{\sigma}}}/2)$ and $\hat{\vec{\sigma}}$ is the vector of Pauli matrices.
The corresponding terms proportional to $\vec{\delta\varphi}$ can be identified and mapped onto generalized Heisenberg model. The expressions for various components of $J^{\alpha\beta}_{ij}$ (Eq.~\eqref{eq:HamS})  are obtained as
\bea
\mathrm{J}^{xx}_{ij}= \frac{\text{T}}{4}\sum_p \text{Tr}_{L,m} \bigl[ \hat{\mathcal{H}}_i , \hat \sigma^{y} \bigl] G_{ij}(i\omega_p)   \bigl[ \hat{\mathcal{H}}_j , \hat \sigma^{y} \bigl] G_{ji} (i\omega_p) \nonumber \\
\mathrm{J}^{xy}_{ij}= -\frac{\text{T}}{4}\sum_p \text{Tr}_{L,m}  \bigl[ \hat{\mathcal{H}}_i , \hat \sigma^{y} \bigl] G_{ij}(i\omega_p)   \bigl[ \hat{\mathcal{H}}_j , \hat \sigma^{x} \bigl] G_{ji} (i\omega_p) \nonumber
\eea
and similar expressions are also for $J\mathrm{J}{ij}^{yy}$ and $\mathrm{J}_{ij}^{yx}$.  The summation is done over the Matsubara frequencies ($\omega_p$) and the trace is over the orbital indices denoted by $m$.  The other components $\mathrm{J}^{xz}_{ij}$, $\mathrm{J}^{zx}_{ij}$, $\mathrm{J}^{yz}_{ij}$, $\mathrm{J}^{zy}_{ij}$ are not of the second order in the tilting angles.  Thus, for M $||$ $z$, only $D_z$ component (Eq.~\eqref{def-int}) can be computed, while $\mathrm{D}_x$ and $\mathrm{D}_y$ are extracted from two additional calculations with the magnetization pointing along $x$ and $y$, respectively. We have used the method described above to calculate the magnetic exchange interactions of YMn$_6$Sn$_6$ as implemented within RSPt code \cite{wills2010full}. 

\subsection{Berry curvature and topological invariant}

The Chern numbers (topological invariants) and Berry curvatures are calculated using a formalism based on Ref.~\cite{fukui}. The Chern number of n'th band over a two-dimensional torus T$^2$ is given by
\begin{equation}
c_n=\frac{1}{2\pi i}\int_{T^2}d^2k\,F_{12}(k),
\label{CheNumCon}
\end{equation}
where the Berry field (Berry curvature)
strength~$F_{12}(k)$ and Berry connection~$A_\mu(k)$ ($\mu=1$, 2) are given by
\begin{eqnarray}
      &&F_{12}(k)=\partial_1A_2(k)-\partial_2A_1(k),
\nonumber\\
&&A_\mu(k)=\braOket{k}{\partial_\mu}{k},
\label{cont-f}
\end{eqnarray}
with $\ket{k}$ being a normalized wave function of the $n$th Bloch band
such that $H(k)\ket{k}=E_n(k)\ket{k}$.

Now we consider a two-dimensional Brillouin zone (BZ) where $\mu=1, 2$ could be $x,y,z$ consisting of discrete lattice points~$k_\ell$ given by 
($\ell=1$, \dots, $N_1N_2$) given by
\begin{equation}
   k_\ell=(k_{j_1},k_{j_2}),\quad
   k_{j_\mu}=\frac{2\pi j_\mu}{q_\mu N_\mu},\quad
   (j_\mu=0,\ldots,N_\mu-1).
\end{equation}
The Berry curvature is defined as 
\begin{eqnarray}
   &&\tilde F_{12}(k_\ell)\equiv
   \ln U_1(k_\ell)U_2(k_\ell+\hat1)U_1(k_\ell+\hat2)^{-1}U_2(k_\ell)^{-1},
\nonumber\\
   &&-\pi<\frac{1}{i}{\tilde F}_{12}(k_\ell)\leq\pi.
\label{FieStr}
\end{eqnarray}
where $\U(1)$ is the scalar product of the wave function of the $n$th
band at two consecutive reciprocal points in the BZ
\begin{equation}
U_\mu(k_\ell)\equiv \braket{k_\ell | k_\ell+\hatmu}/{\cal N}_\mu(k_\ell),
\label{seven}
\end{equation}
with ${\cal N}_\mu(k_\ell)=|\braket{k_\ell | k_\ell+\hatmu}|$.
Finally, a new Chern number for $n$th band is calculated by summing up the imaginary part of the Berry curvature for the discrete points on the BZ and it is defined as
\begin{equation}
   \tilde c_n\equiv\frac{1}{2\pi i}\sum_\ell\tilde F_{12}(k_\ell).
\end{equation}
We used 400$\times$400$\times$1  kmesh to calculate the Chern numbers and Berry curvature for YMn$_6$Sn$_6$  using the above formalism.

\section{Crystal structure of YMn$_6$Sn$_6$}

\begin{figure} [ht] 
\includegraphics[width=0.51\textwidth,angle=0]{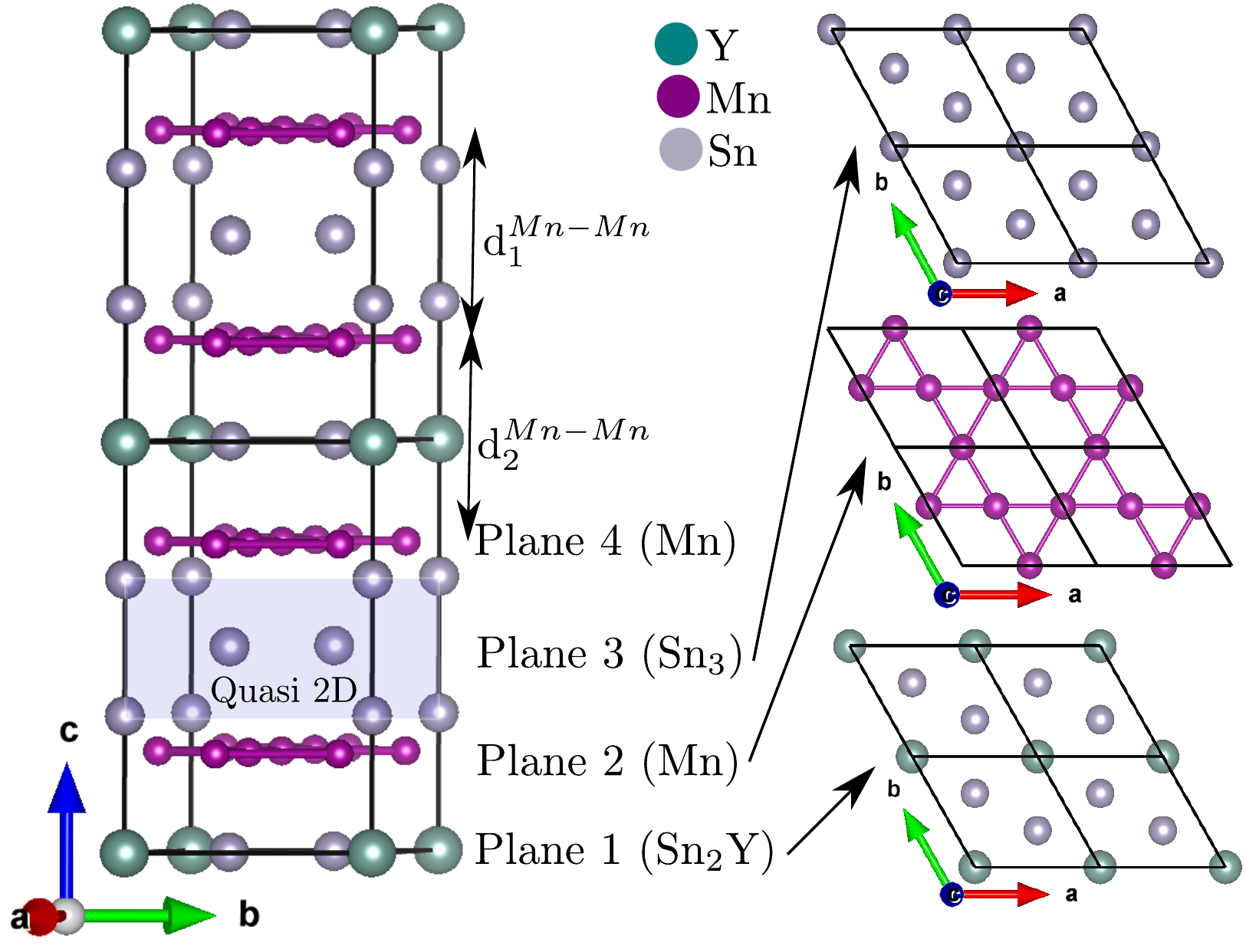} 
\caption{Crystal structure of YMn$_6$Sn$_6$.  The black box corresponds to a single unit cell having four different planes [Sn$_2$Y,  Mn,  quasi 2D Sn$_3$,  Mn] including two different types of Mn kagome planes (plane 2 and plane 4).}
\label{sfig1} 
\end{figure}

\par  We used the experimentally determined structure \cite{zhang2020}.  YMn$_6$Sn$_6$ crystallizes in a centrosymmetric layered structure with space group P6/mmm (No.\,191). The lattice parameters are a = b = 5.536 Å and c = 9.019 Å.  
The structure consists of three two-dimensional (2D) planes and one quasi-2D structure, stacked along the c-axis as shown in Fig.\,\ref{sfig1}.
Of the three 2D planes, two are kagome planes made up of Mn atoms (called plane 2 and 4 in the following).
The third 2D plane is a triangular lattice made up of Sn and Y atoms (called plane 1 in the following)
The quasi-2D structure (called plane 3 in the following) consists of Sn atoms, and has a triangular structure in the ab-plane.
The Mn-planes are spaced by the other two plane types in an alternating fashion.  This means that the Mn planes are spaced with two different distances, $d_1^{\mathrm{Mn-Mn}}$ and $d_2^{\mathrm{Mn-Mn}}$  along the c-axis, with $d_1^{\mathrm{Mn-Mn}}=$~4.536 Å and $d_2^{\mathrm{Mn-Mn}}=$~4.483 Å. The shorter distance is when the spacer layer is plane type 1, whereas plane type 3 is responsible for the slightly longer distance between the Mn planes.   The in-plane ${\mathrm{Mn-Mn}}$  distance is 2.768 Å.  The space group for this type of magnetic kagome crystal structure has 24 point group symmetries, including the $C_2$,  $C_3$, $C_6$ and inversion ($P$) symmetries.

\section{Magnetic interactions in YMn$_6$Sn$_6$}

\par
\begin{figure} [ht] 
\includegraphics[width=0.45\textwidth,angle=0]{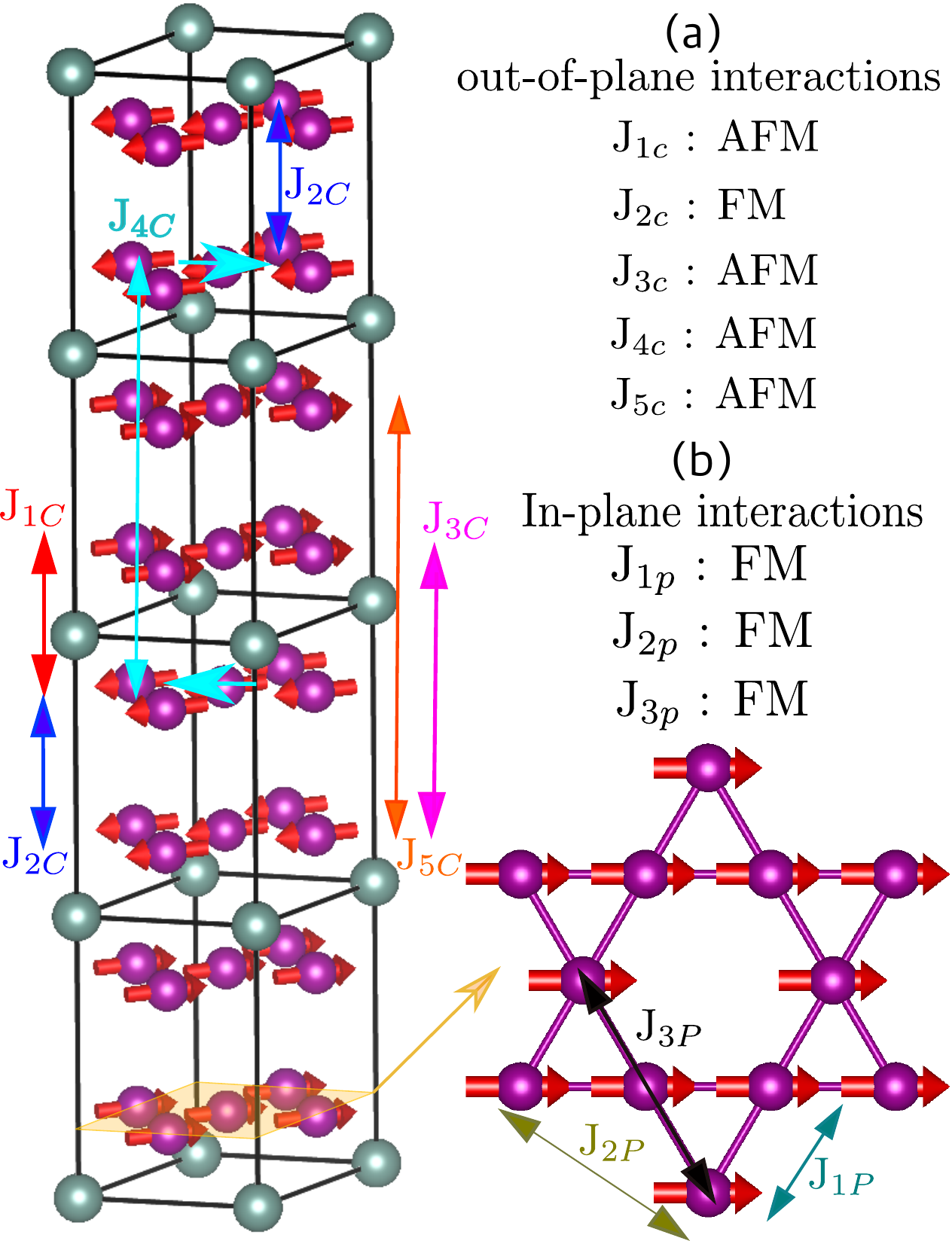} 
\caption{The nearest neighbour magnetic exchange mapping in YMn$_6$Sn$_6$ for (a) out of  plane ($c$-axis) and (b) in-plane ($ab$-plane) directions.}
\label{sfig2} 
\end{figure}

 To describe the exchange interactions properly,  we used 1$\times$1$\times$2 supercell of YMn$_6$Sn$_6$ (see Fig. \ref{sfig1}), where the ${\mathrm{Mn-Mn}}$ across Sn$_2$Y and Sn$_3$ layers have AFM and FM interactions respectively. The calculated magnetic moment using different functionals are presented in the table \ref{tab1}.  The magnetic exchanges in YMn$_6$Sn$_6$ can be categorized into two parts : one is in-plane magnetic exchanges within a kagome plane (J$_{ip}$ with i =1,2,3) and other is magnetic exchanges along the $c$-axis (J$_{ic}$ with i =1,2,3,4,5) as shown in \ref{sfig2}(a)-(b).  The calculated  J$_{ic}$'s (i = 1,2,3,4,5) for YMn$_6$Sn$_6$ from DFT,  DFT with a Hubbard U and DFT+DMFT with a Hubbard U = 2 and 3 eV respectively are presented in a table \ref{tab2}.

\begin{table}[ht]
    \small
    \begin{tabular*}{0.5\textwidth}{ p{4.5cm} p{3.5 cm}  }
    \hline\hline 
       Computational method  & magnetic moment ($\mu_B$) \\ 
           \hline\hline  
 DFT  & 2.36  \\ \\
 DFT+U (U=1 eV)  & 2.34   \\ \\
 DFT+U (U=2 eV)  & 2.26    \\ \\
 DFT+U (U=2.5 eV)  & 2.23    \\ \\
 DFT+U (U=3 eV)  &  2.21  \\ \\
 DFT+U (U=4 eV)  &  2.01  \\ \\
 DFT+DMFT+U (U=2 eV)  & 2.29   \\ \\
  DFT+DMFT+U (U=2.5 eV)  & 2.26   \\ \\
 DFT+DMFT+U (U=3 eV)  & 2.23   \\ 
    \hline 
     \end{tabular*}
     \caption{Magnetic moment of YMn$_6$Sn$_6$ calculated from DFT and DFT+DMFT with inclusion of Hubbard U within LSDA.}
\label{tab1} 
 \end{table}

\begin{table*}[ht]
    \small
    \begin{tabular*}{1.0\textwidth}{ p{6.2cm} p{1.8 cm} p{1.8 cm} p{2.8 cm} p{2.8 cm} p{2.6 cm} }
    \hline\hline 
       Method ($\phi$,$\delta$ from spin dynamics simulation)  & J$_{1c}$ (AFM) & J$_{2c}$ (FM) & J$_{3c}$ (AFM) & J$_{4c}$ (AFM) & J$_{5c}$ (AFM) \\  
         \hline\hline \\ 
 DFT ($\phi=$0, $\delta$=0)  & -11.74 meV  & 8.66 meV   & -1.61 meV   &  -0.37 meV  &  -1.23 meV  \\ \\
 {\rb { DFT+U=2 ($\phi=$54$^{\circ}$, $\delta$=(1-10)$^{\circ}$)}}  &  -8.01 meV  & 7.09 meV  &  -1.21 meV  & -0.49 meV   & -1.07 meV   \\ \\
     {\rb {DFT+U=3 ($\phi=$102$^{\circ}$, $\delta$=(3-11)$^{\circ}$) }} & -4.51  meV  & 3.01  meV  & -0.66  meV  & -0.33  meV   & -0.45  meV   \\ \\
 {\rb { DFT+DMFT+U=2  ($\phi=$58$^{\circ}$, $\delta$=1.62$^{\circ}$)}} & -11.13 meV  & 7.81 meV   & -1.49 meV   &  -0.69 meV  & -1.22 meV   \\
 \\
 {\rb { DFT+DMFT+U=3  ($\phi=$89.3$^{\circ}$, $\delta$=8.9$^{\circ}$)}} & -9.17  meV  & 8.61 meV   &  -1.18  meV   & -0.02 meV  & -0.65  meV   \\ \\
    \hline
Method ($\phi$,$\delta$ from effective model)  & J$_{1c}^{eff}$ (FM) & J$_{2c}^{eff}$ (FM) & J$_{3c}^{eff}$ (FM/AFM) & J$_{4c}^{eff}$ (FM/AFM) & J$_{5c}^{eff}$ (AFM) \\  
         \hline\hline \\ 
 DFT ($\phi=$0, $\delta$=0)  & 1.12  meV  & 18.11 meV   & 0.702 meV (FM)   &  0.21 meV (FM)  &  -1.63 meV  \\ \\
 DFT+DMFT+U=1 ($\phi=$0, $\delta$=0)  & 1.38  meV  & 15.26  meV  & 0.45  meV (FM)  & -0.08 meV  (AFM)   & -1.32 meV    \\ \\
 {\rb {DFT+DMFT+U=2  ($\phi=$58.15$^{\circ}$, $\delta$=9.1$^{\circ}$)}} & 2.49 meV  & 15.39 meV & 0.13 meV (FM) &  -0.54 meV (AFM)  & -1.16 meV  \\ \\
  {\rb {DFT+DMFT+U=2.5 ($\phi=$77.7$^{\circ}$, $\delta$=16.2$^{\circ}$)}} & 3.99  meV  & 16.44  meV & -0.34 meV (AFM) & -0.89  meV (AFM)  & -1.67 meV   \\ \\
  {\rb {DFT+DMFT+U=3  ($\phi=$91.07$^{\circ}$, $\delta$=14.77$^{\circ}$)}} & 4.08  meV  & 16.57  meV & -1.32 meV (AFM) & -1.62  meV (AFM)  & -0.64 meV   \\ \\
DFT+DMFT+U=3.5  ($\phi=$0, $\delta$=0) & 4.11  meV  & 17.12 meV & 0.31  meV (FM) & 0.61  meV (FM)  &  -0.73 meV   \\ \\
 DFT+DMFT+U=4  ($\phi=$0, $\delta$=0) &  4.68 meV  & 18.38 meV & 1.21  meV (FM) & 0.65  meV (FM)  & -0.71 meV   \\ \\  
 \hline \\
 DFT+U=1 ($\phi=$0, $\delta$=0)  & 1.50  meV  & 16.66  meV  & 0.55  meV (FM)  & -0.06 meV  (AFM)   & -1.59  meV    \\ \\  
 {\rb {DFT+U=2 ($\phi=$57.56$^{\circ}$, $\delta$=8.38$^{\circ}$) }} &  1.48 meV  & 16.01 meV  &  0.43 meV (FM)  & -0.22 meV (AFM)   & -1.56 meV  \\ \\
 {\rb {DFT+U=2.5 ($\phi=$95.27$^{\circ}$, $\delta$=11.01$^{\circ}$)}} & 1.37  meV  & 15.76  meV & 0.15 meV (FM) & -0.68  meV (AFM)  & -1.72 meV   \\ \\
{\rb { DFT+U=3 ($\phi=$108.85$^{\circ}$, $\delta$=10.69$^{\circ}$)}}  & 1.79 meV  & 12.71  meV  & -0.56  meV (AFM)   & -0.78 meV (AFM)   &  -0.95 meV   \\ \\
DFT+U=3.5 ($\phi=$0, $\delta$=0)  & 1.37 meV  & 13.42  meV  & 0.17  meV (FM)   & 0.67 meV (FM)   &  -0.89 meV
\\ \\
     \hline    \hline  
     \end{tabular*}
      \caption{Magnetic exchange interactions along the $c$-axis (J$_{ic}$ with i =1,2,3,4, 5) and effective magnetic interactions within Mn planes along the $c$-axis (J$_{ic}^{eff}$ with i =1,2,3,4, 5)  calculated from DFT and DFT+DMFT with inclusion of Hubbard U within LSDA using 1$\times$1$\times$2 supercell of YMn$_6$Sn$_6$.  The effective model is taken from Hamiltonian \ref{model-ham} in Sec. \ref{effc-ham}(B) to calculate $\phi$,$\delta$ for second case which agreed well with our simulated results.  Here we used the convention of positive (negative) J$_{ij}$'s as FM (AFM) respectively throughout the text.}
\label{tab2}     
 \end{table*}

\section{Spin texture of YMn$_6$Sn$_6$}
\label{effc-ham}

\subsection{Spin texture from spin dynamics simulation}

\par Table \ref{tab2} shows pitch angle $\phi$ and the angle between two spirals from spin dynamics simulation (described in the main text) using full magnetic interactions parameters calculated with LSDA,  LSDA+U and LSDA+DMFT with Hubbard U = 2, 3 eV respectively.  We got magnetic ground state as a combination of two incommensurate spin spirals with a pitch angle $\phi$ of 54$^{\circ}$ (102$^{\circ}$) originating from two different Mn kagome planes of YMn$_6$Sn$_6$ from LSDA+U with Hubbard U = 2 (3) eV respectively. The angle between two incommensurate spin spirals $\delta$ is also varying from (1-10)$^{\circ}$ ((3-11)$^{\circ}$) from LSDA+U with Hubbard U = 2 (3) eV respectively.   We got the $\phi$ = 58$^{\circ}$ (89.3$^{\circ}$) and $\delta$ = 1.62$^{\circ}$ (8.9$^{\circ}$) from LSDA+DMFT with Hubbard U = 2 (3) eV respectively.  We compare spin spiral wave vector q, pitch angle $\phi$ and angle between two incommensurate spin spirals $\delta$ of our magnetic ground state calculated from DFT+DMFT with a Hubbard  U = 2, 3 eV with other experimental reports at low temperatures which are presented in the table \ref{tab3}.  Simulated $\phi$ and $\delta$ from LSDA+DMFT with Hubbard U = 3 eV matches well with experiment as shown in table \ref{tab3}. The magnetic exchange interactions and spin spiral in YMn$_6$Sn$_6$ are very sensitive to strucrural parameters, temperatures and magnetic fields.

\begin{table*}[ht]
    \small
    \begin{tabular*}{1.05\textwidth}{ p{7.5cm} p{4.0 cm} p {2.4 cm} p{4.0 cm} }
    \hline\hline  \\ 
       Experimental report  &  spin spiral wave vector & pitch angle $\phi$ & angle between two spirals $\delta$ \\  \\ 
         \hline\hline  \\ 
{\textcolor{magenta}{{Ghimire et al. }}} \cite{ghimire2020} & 0.25 & 90$^{\circ}$ & 20$^{\circ}$ \\
      d$\mathrm{_{[{Sn}_2Y]}} >$  d$\mathrm{_{[{Sn}_3]}}$   &  &  &    \\ \\
          J$_{1c}$ (J$\mathrm{_{[{Sn}_3]}}$) : FM,  J$_{2c}$ (J$\mathrm{_{[{Sn}_2Y]}}$) : AFM,  J$_{3c}$ : FM &  &  &    \\ \\   \\ 
{\textcolor{magenta}{{Xu et al. }}} \cite{xu2021} & 0.266 & 95.6$^{\circ}$ & -  \\ 
                 d$\mathrm{_{[{Sn}_2Y]}} >$  d$\mathrm{_{[{Sn}_3]}}$      &  &  &    \\ \\
          J$_{1c}$ (J$\mathrm{_{[{Sn}_3]}}$) : FM,  J$_{2c}$ (J$\mathrm{_{[{Sn}_2Y]}}$) : FM,  J$_{3c}$ : AFM &  &  &    \\ \\   \\ 
{\textcolor{magenta}{{Rosenfeld et al. }}} \cite{rosenfeld2008} & q$_1$ = 0.2796  & 101$^{\circ}$ & 11$^{\circ}$  \\ 
                d$\mathrm{_{[{Sn}_2Y]}} >$  d$\mathrm{_{[{Sn}_3]}}$   & and q$_2$ = 0.2545  &  &     \\ \\
          J$_{1c}$ (J$\mathrm{_{[{Sn}_3]}}$) : FM,  J$_{2c}$ (J$\mathrm{_{[{Sn}_2Y]}}$) : FM,  J$_{3c}$ : AFM &  &  &    \\ \\   \\ 
{\textcolor{magenta}{{Zhang et al. }}} \cite{zhang2020} & 0.266 & 95.6$^{\circ}$ & -  \\  
         d$\mathrm{_{[{Sn}_2Y]}} <$  d$\mathrm{_{[{Sn}_3]}}$   &  &  &     \\ \\
          J$_{1c}$ (J$\mathrm{_{[{Sn}_2Y]}}$) : AFM,  J$_{2c}$ (J$\mathrm{_{[{Sn}_3]}}$) : FM,  J$_{3c}$ : - &  &  &    \\ \\   \\      
{\textcolor{magenta}{{Our study : simulation with U = 2 eV}}} & 0.1615 & $\sim$ 58$^{\circ}$ ($\pm 0.2$) & $\sim$ 1.62$^{\circ}$ ($\pm 0.01$)  \\ 
         d$\mathrm{_{[{Sn}_2Y]}} <$  d$\mathrm{_{[{Sn}_3]}}$              &  &  &    \\  \\
          J$_{1c}$ (J$\mathrm{_{[{Sn}_2Y]}}$) : AFM,  J$_{2c}$ (J$\mathrm{_{[{Sn}_3]}}$) : FM,  J$_{3c}$ : AFM &  &  &    \\ \\   \\
 {\textcolor{magenta}{{Our study : simulation with U = 3 eV}}} & 0.248 & $\sim$ 89.5$^{\circ}$ ($\pm 0.5$) & $\sim$ 8.9$^{\circ}$ ($\pm 0.06$)  \\ 
         d$\mathrm{_{[{Sn}_2Y]}} <$  d$\mathrm{_{[{Sn}_3]}}$              &  &  &    \\  \\
          J$_{1c}$ (J$\mathrm{_{[{Sn}_2Y]}}$) : AFM,  J$_{2c}$ (J$\mathrm{_{[{Sn}_3]}}$) : FM,  J$_{3c}$ : AFM &  &  &    \\ \\   \\         
{\textcolor{magenta}{{Our study : effective model from Sec. \ref{effc-ham}}, U = 2 eV }} &  & $\sim$ 58.15$^{\circ}$ & $\sim$  9.1$^{\circ}$  \\ 
        d$\mathrm{_{[{Sn}_2Y]}} <$  d$\mathrm{_{[{Sn}_3]}}$              &  &  &    \\  \\
          J$_{1c}^{eff}$ (J$\mathrm{_{[{Sn}_2Y]}}$) : FM,  J$_{2c}^{eff}$ (J$\mathrm{_{[{Sn}_3]}}$) : FM,  J$_{3c}^{eff}$ : FM &  &  &    \\ \\   \\  
{\textcolor{magenta}{{Our study : effective model from Sec. \ref{effc-ham}}, U = 2.5 eV }} &  & $\sim$ 77.7$^{\circ}$ & $\sim$  16.2$^{\circ}$  \\ 
        d$\mathrm{_{[{Sn}_2Y]}} <$  d$\mathrm{_{[{Sn}_3]}}$              &  &  &    \\  \\
          J$_{1c}^{eff}$ (J$\mathrm{_{[{Sn}_2Y]}}$) : FM,  J$_{2c}^{eff}$ (J$\mathrm{_{[{Sn}_3]}}$) : FM,  J$_{3c}^{eff}$ : AFM &  &  &    \\ \\   \\            
       {\textcolor{magenta}{{Our study : effective model from Sec. \ref{effc-ham}}, U = 3 eV }} &  & $\sim$ 91.07$^{\circ}$ & $\sim$  14.47$^{\circ}$  \\ 
               d$\mathrm{_{[{Sn}_2Y]}} <$  d$\mathrm{_{[{Sn}_3]}}$              &  &  &    \\  \\
 J$_{1c}^{eff}$ (J$\mathrm{_{[{Sn}_2Y]}}$) : FM,  J$_{2c}^{eff}$ (J$\mathrm{_{[{Sn}_3]}}$) : FM,  J$_{3c}^{eff}$ : AFM &  &  &    \\ \\   \\               
    \hline 
     \end{tabular*}
 \caption{Spin spiral wave vector q,  pitch angle $\phi$ and angle between two spin spirals $\delta$ in YMn$_6$Sn$_6$ from different experimental studies at low temperatures.  Magnetic exchange interactions and ground state are very much sensitive to structural parameters.  d$\mathrm{_{[{Sn}_3]}}$ and d$_\mathrm{{[{Sn}_2Y]}}$ are the ${\mathrm{Mn-Mn}}$ layers distance separated by $\mathrm{{[{Sn}_3]}}$ and $\mathrm{{[{Sn}_2Y]}}$ layers respectively as described in Fig.\ref{sfig1}.  J$\mathrm{_{[{Sn}_3]}}$ and J$_\mathrm{{[{Sn}_2Y]}}$ are the ${\mathrm{Mn-Mn}}$ layers exchange interactions separated by $\mathrm{{[{Sn}_3]}}$ and $\mathrm{{[{Sn}_2Y]}}$ layers respectively. }
\label{tab3}    
 \end{table*}

\begin{figure} [ht] 
\centering
\includegraphics[width=0.45\textwidth,angle=0]{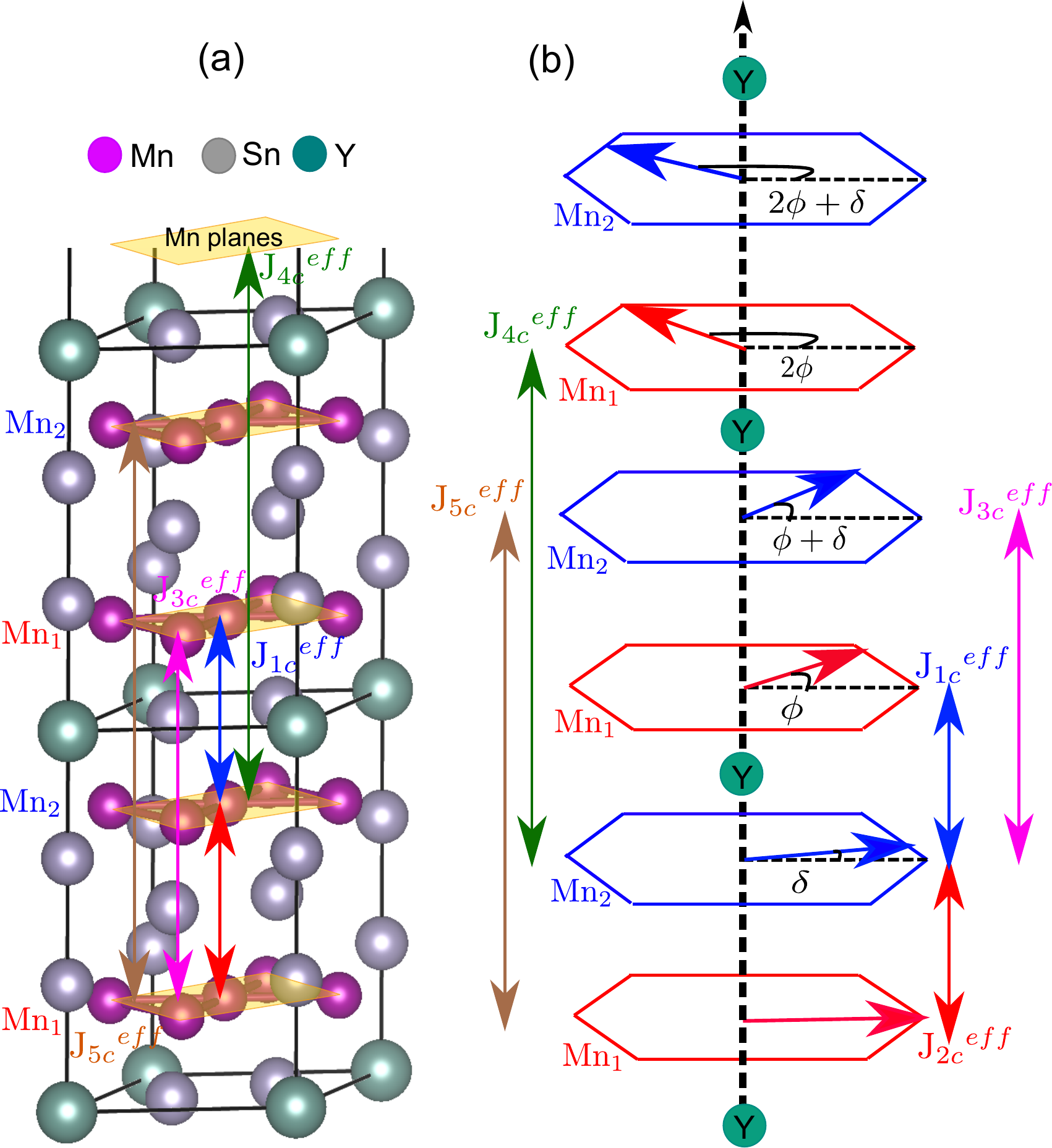} 
\caption{(a) Effective magnetic interactions within Mn planes along the $c$-axis (J$_{ic}^{eff}$ with i =1,2,3,4, 5) in  YMn$_6$Sn$_6$. (b) Effective model of two spin spirals with five effective magnetic exchanges along $c$-axis J$_{ic}^{eff}$ with i =1,2,3,4, 5.  Here $\phi$ is the pitch angle for both spirals and $\delta$ is the angle between two spin spirals.}
\label{sfig3} 
\end{figure}

\subsection{Effective magnetic Hamiltonian for spin spiral of YMn$_6$Sn$_6$}
We consider an effective Hamiltonian to describe the two incommensurate spin spirals in  YMn$_6$Sn$_6$ given by :
\begin{eqnarray}
{\mathrm{H}}_{ic}^{eff} = - {\mathrm{J}}_{1c}^{eff} {\cos(\phi - \delta)} - {\mathrm{J}}_{2c}^{eff} {\cos\delta} - 2{\mathrm{J}}_{3c}^{eff} {\cos\phi} \nonumber \\ - {\mathrm{J}}_{4c}^{eff} {\cos(2\phi - \delta)} - {\mathrm{J}}_{5c}^{eff} {\cos(\phi + \delta)} 
\label{model-ham}
\end{eqnarray}
where H$_{ic}^{eff}$ is the effective model for two spin spiral with 5 magnetic effective magnetic exchanges along $c$-axis (J$_{ic}^{eff}$ for i=1,2,3,4,5) extended from Rosenfeld et al.  model with 3 magnetic effective magnetic exchanges along $c$-axis \cite{rosenfeld2008} (see Fig.\ref{sfig3}). Here $\phi$ is the pitch angle for both spirals and $\delta$ is the angle between two spin spirals. The effective magnetic exchanges J$_{ic}^{eff}$ are calculated by summing over all out of plane magnetic exchanges within a plane. For example : J$_{1c}^{eff}$ = (J$_2$+4$\times$J$_5$+4$\times$J$_8$),  J$_{2c}^{eff}$ = (J$_3$+4$\times$J$_6$+4$\times$J$_9$) and so on as shown in Fig.1(b) in the main text and Fig.\ref{sfig3} (a).  The effective magnetic exchange interactions J$_{ic}^{eff}$ for i=1,2,3,4,5 calculated from different methods are shown in table \ref{tab2}. The calculated $\phi$ and $\delta$ using effective magnetic exchanges between Mn planes along $c$-axis (J$_{ic}^{eff}$ for i=1,2,3,4,5) are presented in table \ref{tab2} for three different methods like LSDA,  LSDA+U (with U = 1,  2,  2.5, 3,  3.5 eV) and LSDA+DMFT with a Hubbard U = 1,  2,  2.5, 3,  3.5, 4  eV.  YMn$_6$Sn$_6$ gives a spin spiral phase with Hubbard U = (2-3) eV and remains in FM phase for others U values from both LSDA+U and LSDA+DMFT with U as shown in table \ref{tab2}.  The calculated values are matches well with simulated spin spirals within UppASD \cite{uppasd} for three different methods presented in table \ref{tab2}.

\par
\begin{figure*} [ht] 
\centering
\includegraphics[width=1.0\textwidth,angle=0]{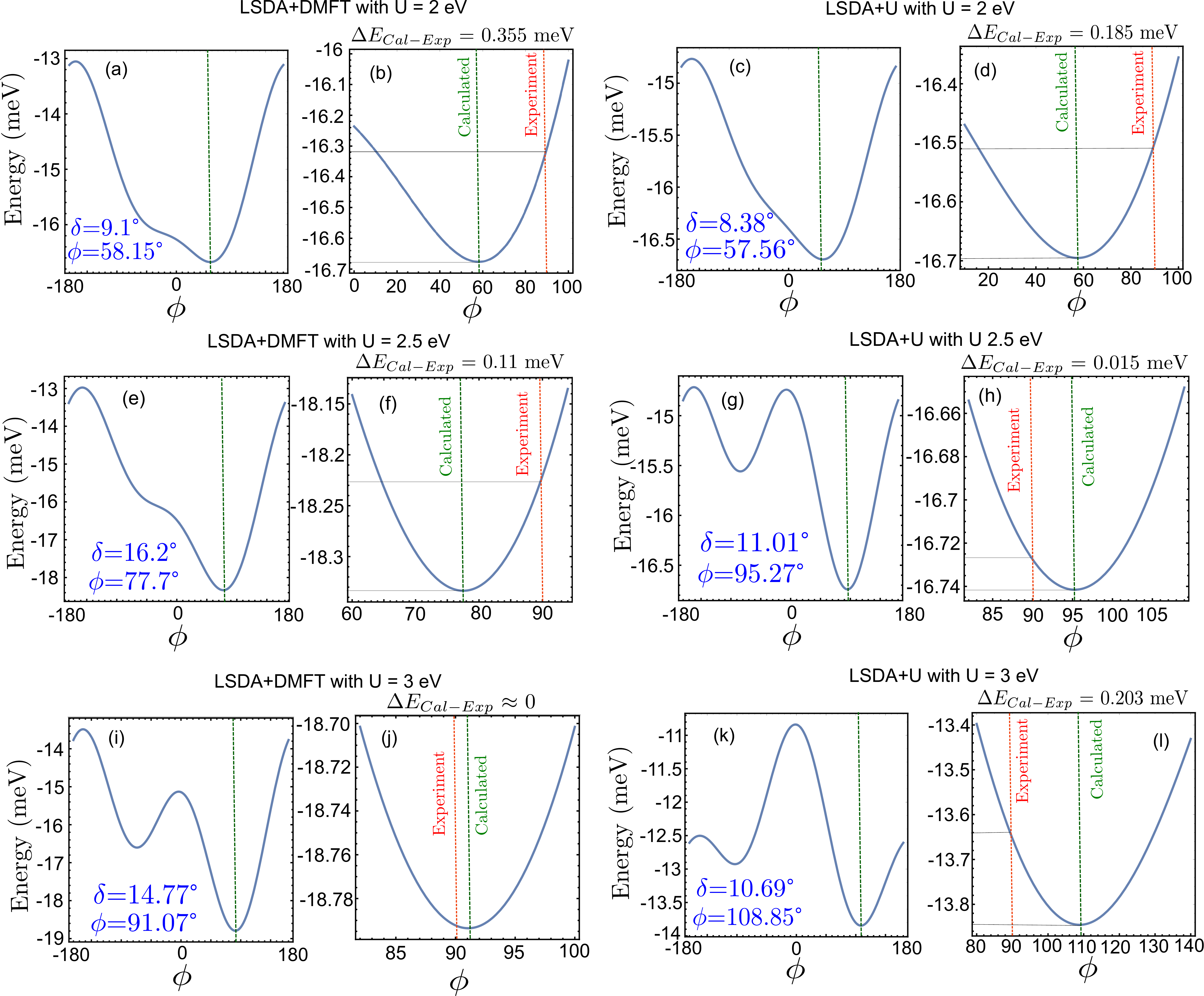} 
\caption{Calculated variation of energy with the pitch angle from (a)-(b) LSDA+DMFT and (c)-(d) LSDA+U with  U = 2 eV,  (e)-(f)  LSDA+DMFT and (g)-(h) LSDA+U with  U = 2.5 eV,  (i)-(j) LSDA+DMFT and (k)-(l) LSDA+U with  U = 3 eV using the model Hamiltonian in  \ref{model-ham} (Sec. \ref{effc-ham}(B)) with calculated $\delta$. The calculated pitch angles are compared with experimental report \cite{ghimire2020}.}
\label{sfig4} 
\end{figure*}

\begin{figure*} [ht] 
\centering
\includegraphics[width=0.9\textwidth,angle=0]{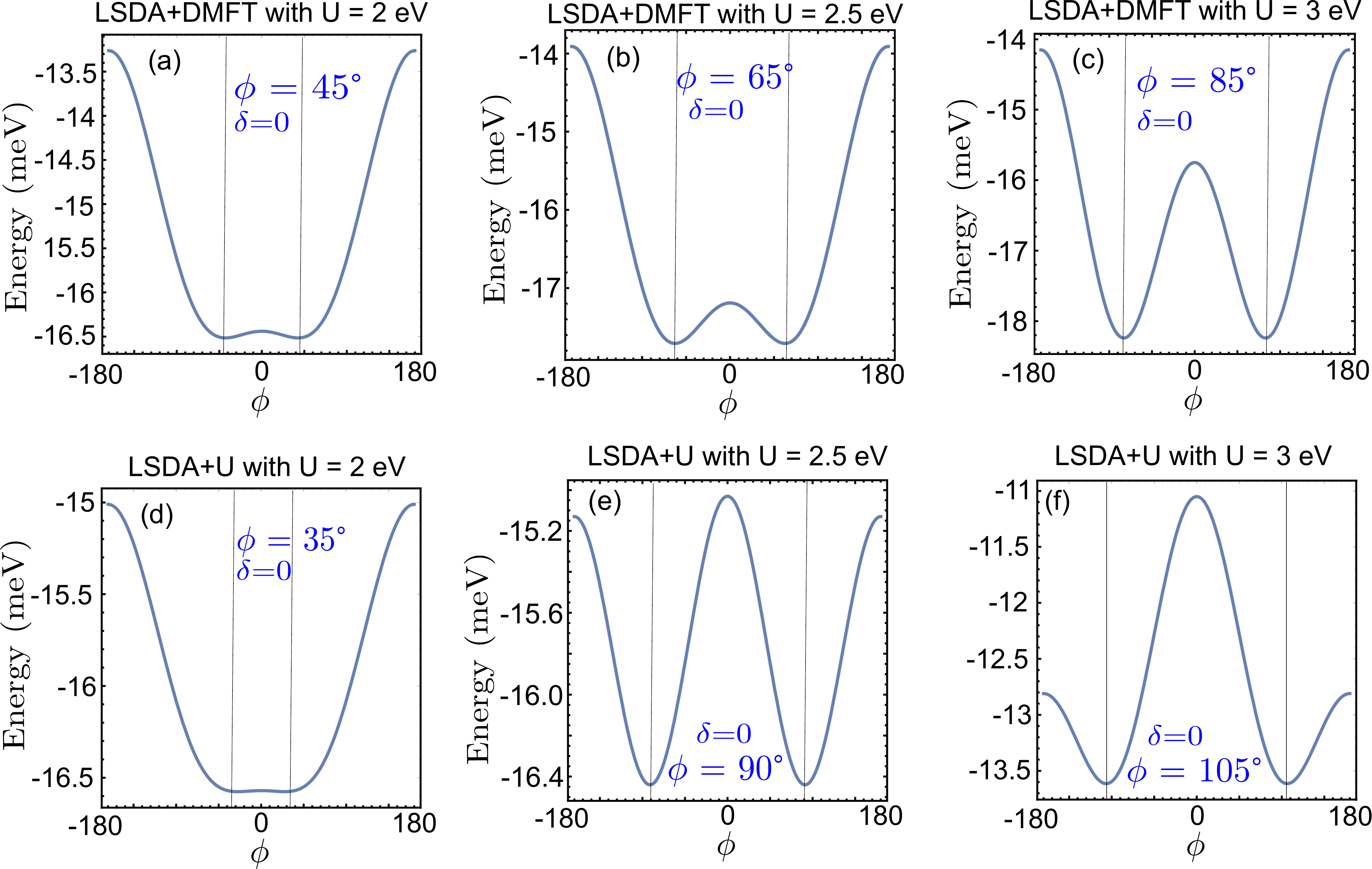} 
\caption{Calculated variation of energy with the pitch angle from (a) LSDA+DMFT with U = 2 eV and (b) LSDA+U with  U = 2 eV,  (c) LSDA+U with  U = 3 eV using the model Hamiltonian in  \ref{model-ham} (Sec. \ref{effc-ham}) with $\delta$=0.  }
\label{sfig5} 
\end{figure*}

 Figure \ref{sfig4} and \ref{sfig5} represents the energy evolution of pitch angle with calculated $\delta$ and $\delta = 0$ from both LSDA+U and LSDA+DMFT with Hubbard U = 2, 2.5, 3 eV respectively using the effective Hamiltonian \ref{model-ham} (see also table \ref{tab2}).  The energy differences between calculated and experimental (pitch angle $\phi\sim$90$^{\circ}$) spin spirals are 0.355 meV,  0.11 and 0 meV from LSDA+DMFT with U = 2,  2.5, 3 respectively,  whereas the corresponding energy difference is 0.185 meV,  0.015,  0.203 meV from LSDA+U with U = 2, 2.5, 3 eV respectively.   The calculated pitch angle $\phi$ = 91.07$^{\circ}$ for the spin spin spiral matches with the experimental value from LSDA+DMFT with U = 3 eV ($\Delta{E}_{{Cal}-{Exp}}$ = 0), whereas calculated $\phi$ = 95.27$^{\circ}$  for the spin spiral close to the experimental value from LSDA+U with U = 2.5 eV ($\Delta{E}_{{Cal}-{Exp}}$ = 0.015 meV).  Putting $\delta = 0$ i.e angle between two non-equivalent,  incommensurate spin spirals zero,  the asymmetric double well nature of the energy evolution transforms into symmetric double well which indicates the two equivalent,  incommensurate spin spirals in YMn$_6$Sn$_6$ (see Fig.\ref{sfig4} and Fig.\ref{sfig5}).

\section{Magnon band structures of YMn$_6$Sn$_6$}

\par
\begin{figure*} [ht] 
\centering
\includegraphics[width=0.7\textwidth,angle=0]{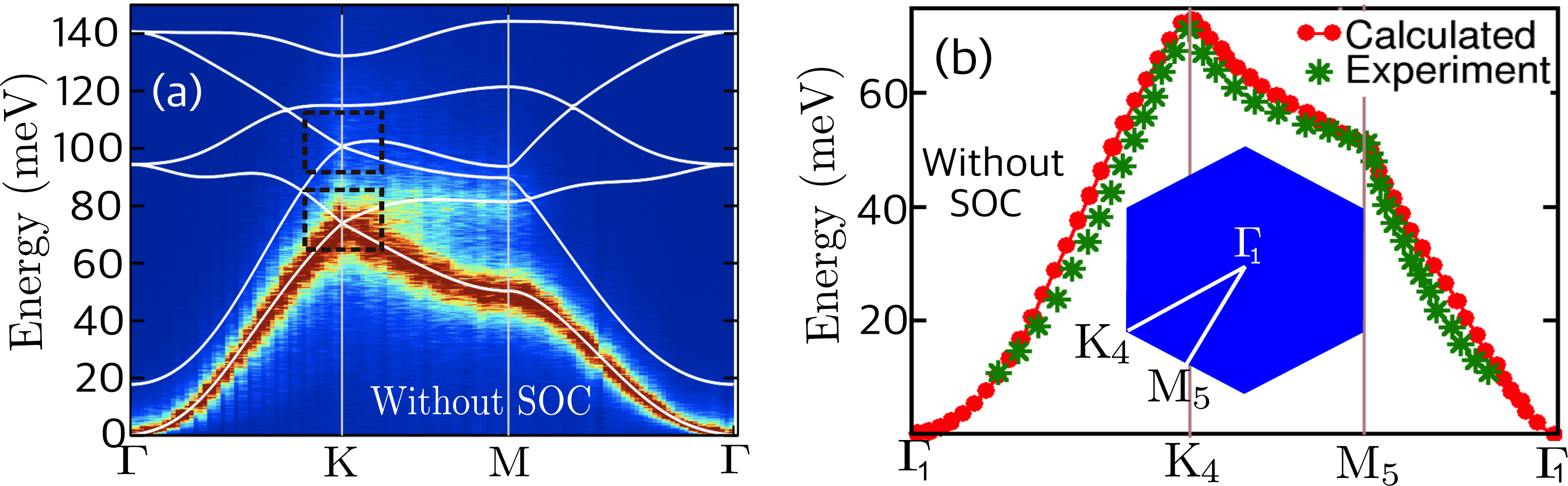} 
\caption{(a) Magnon bands in YMn$_6$Sn$_6$ without SOC along (a) $\Gamma$(0 0  0) - K(1/3  1/3  0) - M(1/2 0  0) - $\Gamma$(0 0 0) and (b) $\Gamma_1$(2 0  0) - K$_4$(5/3  -1/3  0) - M$_5$(2 -1/2  0) - $\Gamma_1$(2 0 0).  The Dirac points are marked by boxes in the left figure and the lowest magnon band is in good agreement with the observed magnon spectra according to the experimental report by Zhang et.al.  \cite{zhang2020}.}
\label{sfig6} 
\end{figure*}

 Figure  \ref{sfig6}(a) shows magnon band structure of YMn$_6$Sn$_6$ without SOC from LSDA+DMFT with U = 2 eV.  Here the calculated magnetic exchanges within a kagome plane are J$_{1p} = 9.52 $ meV, J$_{2p} = 0.76 $ meV respectively and out of plane magnetic exchanges are J$_{1c} = -7.16 $ meV,  J$_{2c} = 4.52 $ meV respectively.  The Dirac band crossing (Dirac magnon) appears at 73 meV at high symmetry K point which is in good agreement with experimental report \cite{zhang2020}.  To compare with calculated  magnon spectra with the experimental study reported by  Zhang  et al. \cite{zhang2020},  we calculated the dynamical structure factor ${\mathrm{S}}({\mathrm{q}},\omega)$ from spin dynamics simulation at T = 5.  This is in good agreement with the experimental measured spectra with LSDA+DMFT with U = 2 eV as shown in \ref{sfig6}(a).  Furthermore,  we compared the lowest magnon band along the path $\Gamma_1$(2 0  0) - K$_4$(5/3  -1/3  0) - M$_5$(2 -1/2  0) - $\Gamma_1$(2 0 0) \cite{zhang2020} which is excellent agreement with the experimental  observation as shown in Fig.\ref{sfig6}(b).  Another Dirac crossing appears at 100 meV.   Both the Dirac crossings open topological gap with inclusion of SOC at that point.

\par
\begin{figure*} [ht] 
\centering
\includegraphics[width=0.9\textwidth,angle=0]{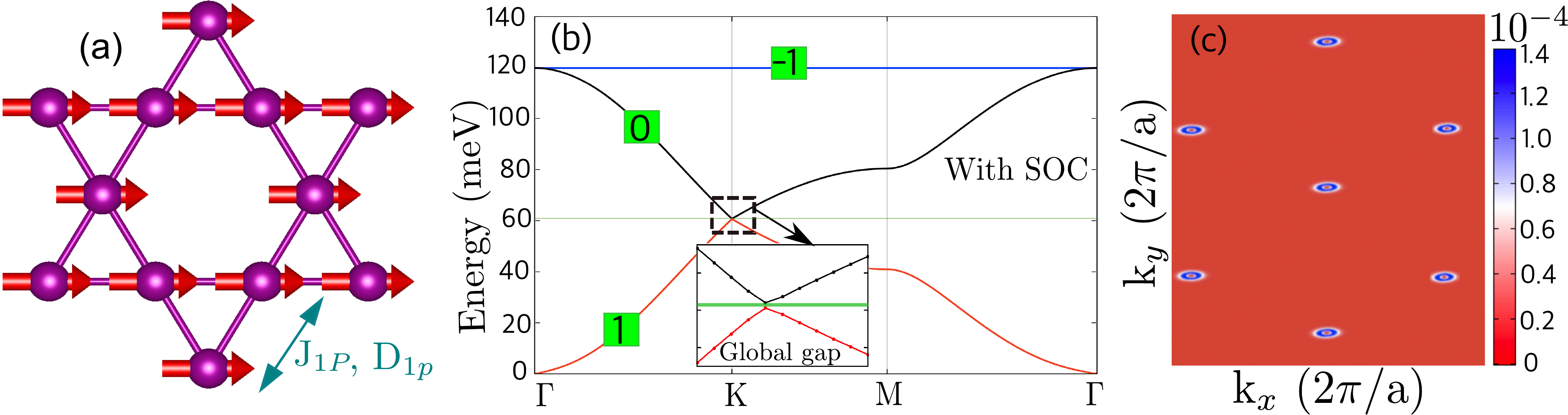} 
\caption{(a) Mn Kagome plane, (b) magnon bands and (c) Berry curvature profile for quasi 2D structure of YMn$_6$Sn$_6$.
}
\label{sfig7} 
\end{figure*}

 We calculated the magnon bands for quasi 2D kagome plane of YMn$_6$Sn$_6$ (as shown in Fig.\ref{sfig7}(a) where  J$_{ij}$'s and D$_{ij}$'s are taken from bulk calculations) to study the effect of in-plane magnetic ($ab$-plane) interactions on magnon topology.  Figure \ref{sfig7}(b) represents the magnon bands from LSDA+DMFT with U = 3 eV with SOC.  Here the in-plane magnetic exchange J$_{1p}$ $\sim$ 4.85 meV. The in-plane and out of plane components of DMIs are ${D_{1p}}^{ \parallel}$ $\sim$ 0.218 meV and ${D_{1p}}^z$ $\sim$  0 respectively.  We got a gapless mode if include only the first nearest neighbour magnetic interactions as $\frac{{D_{1p}}^z}{J_{1p}} \sim 0$.  However, this opens a gap of $\sim 0.6$ meV at high symmetry $K$ point around $\sim 62$ meV with inclusion of ${D_{2p}}^z$ $\sim$ 0.308 meV as shown in Fig.\ref{sfig7}(b).  The Chern numbers of upper, middle and lower magnon bands are $c = -1,  0,  1$ respectively.  Flat bands with nonzero Chern number ($c = -1$) (see Fig.\ref{sfig7}(b)) and Berry curvature profile ( see Fig.\ref{sfig7}(c)) ensure the landmark of topological magnon in incommensurate spin spiral of YMn$_6$Sn$_6$.

\section{Electronic band structures of YMn$_6$Sn$_6$}

\par
\begin{figure*} [ht] 
\includegraphics[width=0.99\textwidth,angle=0]{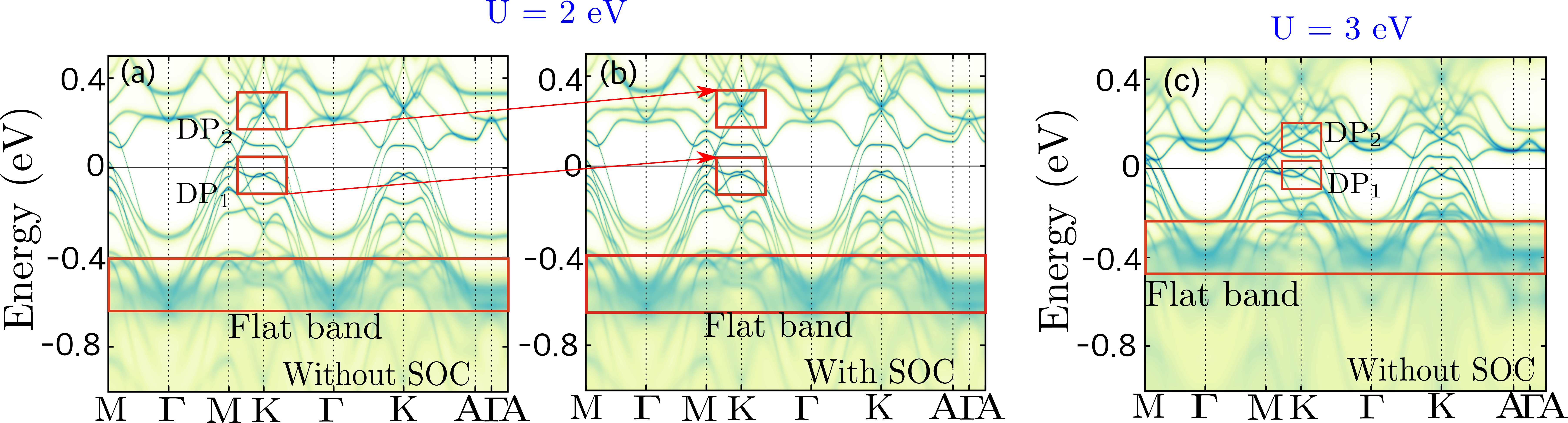} 
\caption{Band structures from LDA+DMFT with (a)-(b) U=2 eV without and with SOC respectively,  and (c) U=3 eV without SOC  where the Dirac points are marked by boxes. }
\label{sfig8} 
\end{figure*}

 We calculated the electronic band structure of YMn$_6$Sn$_6$ from LDA+DMFT with a Hubbard U = 2, 3 eV and a Hund’s coupling J = 0.7 eV.  Figure \ref{sfig8}(a)-(b) show the band structures of YMn$_6$Sn$_6$ without and with SOC along the high symmetry $M-\Gamma-{\mathrm{M}}-{\mathrm{K}}-\Gamma-{\mathrm{K}}-{\mathrm{A}}-\Gamma-{\mathrm{A}}$ directions in the hexagonal Brillouin zone (BZ) with a Hubbard U = 2 eV and a Hund’s coupling J = 0.7 eV on ${\mathrm{Mn-3d}}$ orbitals.  Two Dirac points (DPs) DP$_1$ and DP$_2$ are located at about $\sim 0.038,  0.279$ eV below and above the Fermi level E$_f$ at the high symmetry ${\mathrm{K}}$ point respectively which are in good agreement with experimental report where DPs appeared at $\sim 0.04, 0.3$ eV respectively  \cite{Li2021-bl}.  With inclusion of SOC,  gaps open at high symmetry K point around the DPs.   Another feature is the observation of flat band at about $\sim 0.4$ eV below E$_f$ which exists throughout the whole BZ \cite{Li2021-bl}.  The flat bands are pushed to E$_f$ with further increasing Hubbard interaction to U = 3 eV as shown in Fig. \ref{sfig8}(c).

\end{document}